\documentclass[arguments, twocolumn, 10 point font, single spaced article, twocolappendix]{aastex631}

\usepackage{graphicx}	
\usepackage{amsmath}	
\usepackage{amssymb}	
\usepackage{xspace}
\usepackage[utf8]{inputenc}
\usepackage{ae,aecompl}
\usepackage{comment}
\usepackage{natbib}
\usepackage{float}
\usepackage{graphicx}
\usepackage{amssymb}
\usepackage{rotating,times,pictex,graphicx,latexsym}
\usepackage{color}
\usepackage{amsmath}
\usepackage{threeparttable}
\usepackage{lipsum}
\usepackage{amsmath}
\usepackage[all]{hypcap}
\usepackage[title,titletoc]{appendix}
\usepackage[]{hyperref}
\PassOptionsToPackage{pdfpagelabels=false}{hyperref}
\usepackage[T1]{fontenc}
\usepackage{changepage}

\newcommand{\Ms}{\ensuremath{M_{\odot}}}

\newcommand{\beq}{\begin{equation}}
\newcommand{\eeq}{\end{equation}}

\newcommand{\hii}{\mbox{H~{\sc ii}~}}
\newcommand{\mum}{\,\ensuremath{\mu}m\xspace}

\newcommand{\x}{\,\ensuremath{\times}\,}

\newcommand{\arcs}{\hbox{$^{\prime\prime}$}}

\newcommand{\arcm}{\mbox{$^{\prime}$}}

\newcommand{\lsun}{\mbox{\rm $L_{\odot}$}}

\newcommand{\av}{\mbox{$A_\mathrm{V}$~}}
\newcommand{\ak}{\mbox{$A_\mathrm{K}$~}}

\newcommand{\kms}{\hbox{km~s$^{-1}$}}

\newcommand{\cmq}{\hbox{cm$^{-3}$}}

\newcommand{\nht}{\hbox{N$(\mathrm H_2)$}}

\newcommand{\thco}{\hbox{$^{13}$CO}~}

\newcommand{\eico}{\hbox{C$^{18}$O}~}

\newcommand{\cloud}{{\rm G148.24+00.41}}
\newcommand{\cluster}{{\rm FSR 655}}

\begin{document}

\title[Star Formation in GMC G148.24+00.41]{
\center
\Large{Peering into the heart of the giant molecular cloud G148.24+00.41: A deep near-infrared view of the newly hatched  cluster  FSR 655}}

\author[0000-0001-6515-2863]{Vineet Rawat}
\affiliation{Physical Research Laboratory, Navrangpura, Ahmedabad, Gujarat 380009, India}
\affiliation{Indian Institute of Technology Gandhinagar Palaj, Gandhinagar 382355, India}
\author[0000-0002-9431-6297]{M. R. Samal}
\affiliation{Physical Research Laboratory, Navrangpura, Ahmedabad, Gujarat 380009, India}
\author[0000-0001-9312-3816]{D.K. Ojha}
\affiliation{Department  of  Astronomy  and  Astrophysics,  Tata  Institute  of  Fundamental  Research,  Mumbai  400005, India}
\author[0000-0001-7225-2475]{Brajesh Kumar}
\affiliation{South-Western Institute for Astronomy Research, Yunnan University, Kunming, Yunnan, 650500, People’s Republic of China}
\affiliation{Aryabhatta Research Institute of Observational Sciences, Manora Peak, Nainital 263 002, India}
\author[0000-0001-5731-3057]{Saurabh Sharma}
\affiliation{Aryabhatta Research Institute of Observational Sciences, Manora Peak, Nainital 263 002, India}
\author[0000-0003-4908-4404]{J. Jose}
\affiliation{Indian Institute of Science Education and Research (IISER) Tirupati, Rami Reddy Nagar, Karakambadi Road, Tirupati 517 507, India}
\author[0000-0003-4973-4745]{Ram Sagar}
\affiliation{Indian Institute of Astrophysics, Block II, Koramangala, Bangalore 560034, India}
\author[0000-0002-6740-7425]{R. K. Yadav}
\affiliation{National Astronomical Research Institute of Thailand (NARIT), Sirindhorn AstroPark, 260 Moo 4, T. Donkaew, A. Maerim, Chiangmai 50180, Thailand}

\begin{abstract}
We present a detailed near-infrared study of an embedded cluster located in the hub of the giant molecular cloud \cloud~of mass $\sim$10$^5$ \Ms, with the  TANSPEC instrument mounted on the 3.6 m Devasthal Optical Telescope. The hub is located near the geometric center of the cloud and represents its most massive clump. 
 We studied the central 2 pc $\times$ 2 pc area of the hub with 5$\sigma$ limiting magnitudes of 20.5, 20.1, and 18.6 mag in the $J$, $H$, and $K_s$ bands, respectively. 
Using the $K_s$-band luminosity function and comparing it with the synthetic clusters, we obtained the age of the cluster as $\sim$0.5 Myr, which was found to corroborate well with the visual extinction versus the age of nearby embedded clusters.  
We find that the present mass of the cluster is around $\sim$180 \Ms, and the cluster is currently forming stars at a rate of $\sim$330 \Ms~Myr$^{-1}$, with an efficiency of $\sim$20\%. 
The cluster is connected to an extended gas reservoir through a filamentary network; thus, we hypothesize that the cluster has the potential to become a richer cluster in a few Myr of time.

\end{abstract}



\section{Introduction}
\label{int}

It is believed that the majority of the stars, if not all, form in a clustered environment. The crowded environment in which stars form determines the properties of stars themselves – the initial mass function (IMF), stellar multiplicity distributions, circumstellar disks, and probably their planetary properties as well. 

However, the formation and subsequent evolution of stellar clusters as bound entities remain enigmatic \citep{lada03, kru19}. 
In theories of star cluster formation, it is still debated whether they form monolithically in a single gravitational collapse event or through a hierarchical process involving gas accretion onto protoclusters while stars form concurrently \citep[see reviews by][]{long14,kru20,kra20}. The flow-driven models, such as global hierarchical collapse \citep[GHC;][]{vaz19} and inertial-inflow model \citep[I2;][]{pad20} propose that the large-scale converging flows within molecular clouds facilitate the formation of filamentary structures. These structures act like conveyor belts that transport gaseous matter and locally formed stars from the extended environment toward the dense clumps of the hub, which, by accreting matter from the surroundings, will become massive and likely form massive clusters \citep{vaz17,vaz19,pad20}. 
Observations have identified filament-converging configurations in clouds, fostering cluster formation at junction points known as hub-filament systems, where matter is found to be funneled via converging flows \citep{mye09, pere14, tre19, kum20, Ma_2023, liu23}.

Although the aforementioned mechanism seems to be a viable way of producing star clusters, the rapid stellar feedback from the newly formed stars can significantly affect the star formation rate (SFR) of the cluster.  
The feedback from young massive stars can inhibit the gas collapse and isolate the cluster-forming clump from the cloud. The low star formation efficiency (SFE) within the cluster-forming region can also impact the emergence of a rich and bound cluster.
Observations and simulations suggest that the embedded phase of star clusters lies somewhat between 1 to 3 Myr depending upon their mass, and bound clusters arise only from regions where SFE is higher than 20\%$-$30\% \citep{lada03,kru19}. In addition, it is also suggested that the primordial structure and density profile of the gas 
also play a decisive role in the formation of massive stars and associated clusters \cite[e.g.][]{Bonn06, Parker_2014, chen21}. Thus, the prerequisite condition to improve our understanding of the formation of intermediate to massive star clusters is to investigate a sample of young clusters of different ages and masses that have recently formed in massive clouds. 
Notably, investigating massive bound clouds featuring hub-filamentary configurations with the hub positioned near the cloud's geometric center is of particular interest. Such clouds offer a likelihood scenario, where the cloud collapses and may form a massive cluster via multiple filamentary flows over an extended period of time. 

In a recent work, \cite{raw23} characterized and investigated the massive cloud \cloud~in order to find out its cluster formation potential and mechanism(s) by which an eventual cluster may emerge. \cite{raw23} found that \cloud~(distance $\sim$3.4 kpc) is a bound, massive (mass $\sim$10$^5$ \Ms), and cold (dust temperature $\sim$14.5 K) giant molecular cloud (GMC) of radius $\sim$26 pc, with a hub-filamentary morphology and it shows the signatures of global hierarchical collapse. Based on CO (1$-$0) isotopologue observations (spatial resolution $\sim$52\arcsec and velocity resolution $\sim$0.17 \kms), \cite{raw24a} found that the cloud is host to a massive
clump-C1 (i.e., the clump with ID C1 and mass $\sim$2100 \Ms) near its geometric center. They also found that the C1 clump lies at the nexus of several large-scale (14$-$38 pc) filamentary structures that are combinedly fueling the clump with cold gaseous matter. Using dust continuum observations at 850\mum, \cite{raw24b} further resolved the C1 clump into multiple substructures and found that the central region of the clump, named CC region in their work, is magnetically supercritical with a magnetic field strength of $\sim$24 $\mu$G. Using $\it{Spitzer}$ near-infrared (NIR) images, \cite{raw23} discussed that a cluster had formed in the clump. This cluster appears to be extremely young, as it is barely visible in optical images, 
while the clustering of point sources is clearly seen in the Two Micron All Sky Survey (2MASS) and $\it{Spitzer}$ images. \cite{fro07}, in their search for embedded clusters using 2MASS images, have identified this
cluster as FSR 655. \cite{raw23} found that despite hosting a highly luminous young stellar object (YSO) with a luminosity in the range of 1900$-$4000 \lsun, no radio emission has yet been detected in the clump at 6.7 GHz, implying that an \hii region is not yet developed. Thus, the hub is a potential target for understanding the early formation and evolution of a cluster that is embedded in a large reservoir of gas and dust. Despite its importance, so far, no detailed attention has been paid to characterize the cluster based on deep infrared observations. Based on the $\it{Spitzer}$ and Wide-field Infrared Survey Explorer catalogs, only a few YSOs have been identified in the literature \citep{Winston_2020}. The molecular hydrogen column density (\nht) in the direction of the cluster was found to be in the range of $\sim$2$-$3 $\times$ 10$^{22}$ cm$^{-2}$ \citep{raw23}, implying a high visual extinction (\av $\sim$20$-$30 mag) for cluster members to be detected in optical and shallow NIR bands \citep[i.e., 10$\sigma$ limiting sensitivity of $\sim$14.3 mag in the $K_s$-band;][]{Skrut_2006}. 

In this work, we conduct a deep NIR analysis of the cluster, using the data obtained with the newly installed 3.6 m Devasthal Optical Telescope (DOT), complemented by catalogs from the $\it{Spitzer}$ observations. We aim to improve the understanding of the current status of the cluster in terms of its evolutionary stage, mass distribution, star-formation efficiency and rate, and
likely fate in the context of massive cluster formation. 
This work is organized as follows: observations, data reduction, and archival data sets are presented in Section \ref{sec_obs}. The results are presented and discussed in section \ref{analysis}. In section \ref{summary_con}, we summarize our work with concluding remarks.

\section{Observations and Data Sets}
\label{sec_obs}

\subsection{Near-infrared Observations}


\begin{figure}
    \centering
    \includegraphics[width=4.1cm]{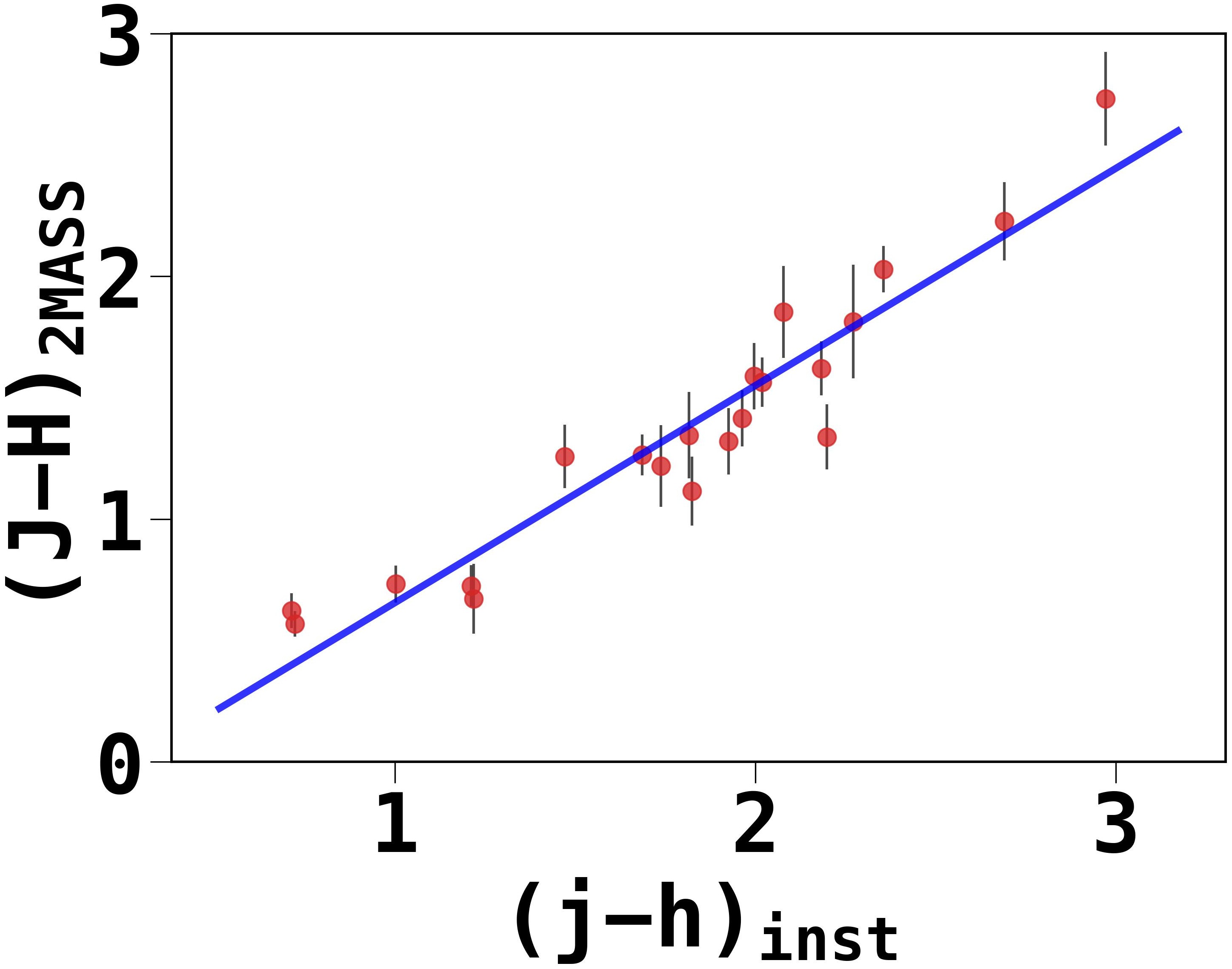}    
     \includegraphics[width=4.1cm]{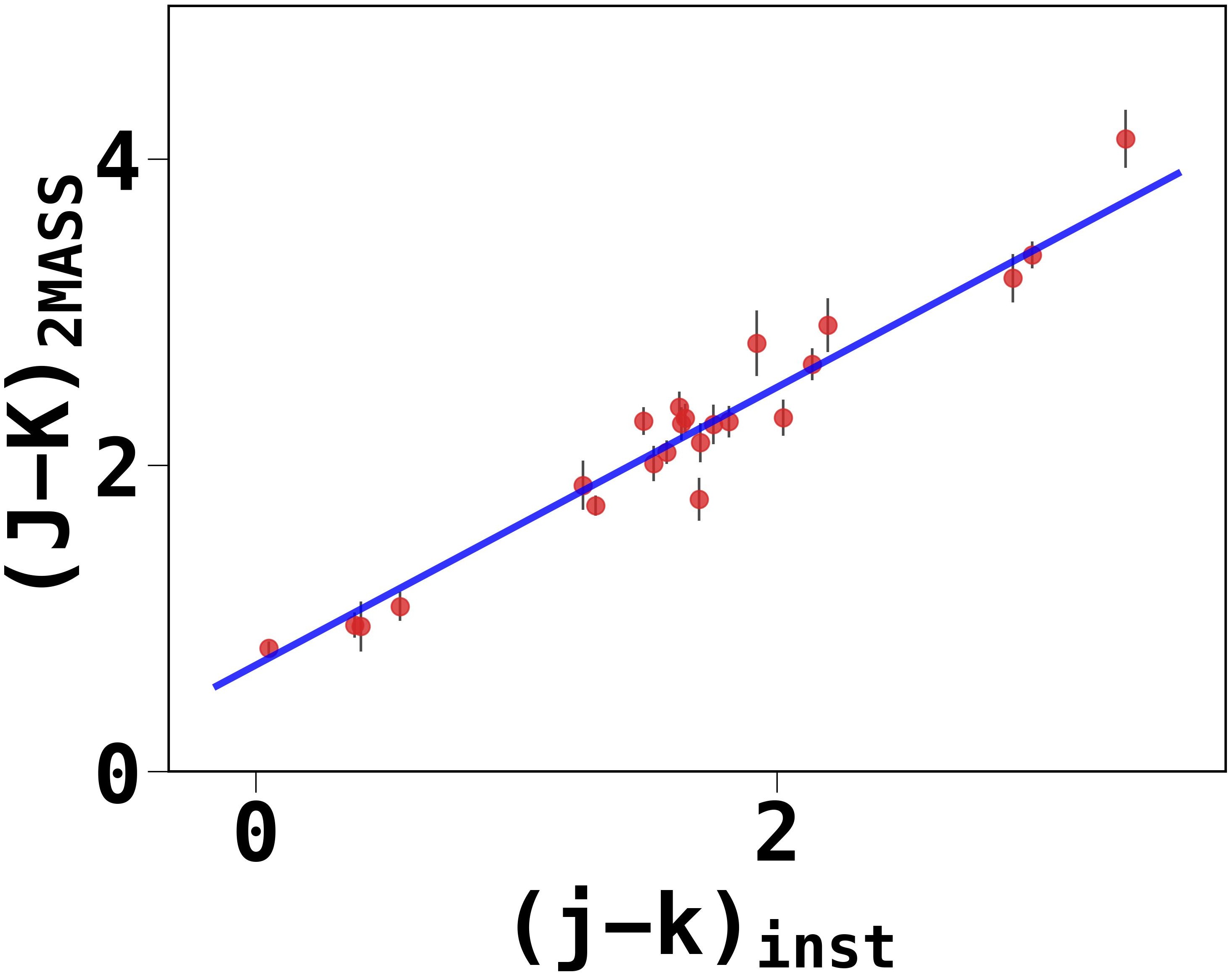}
     \includegraphics[width=4.1cm]{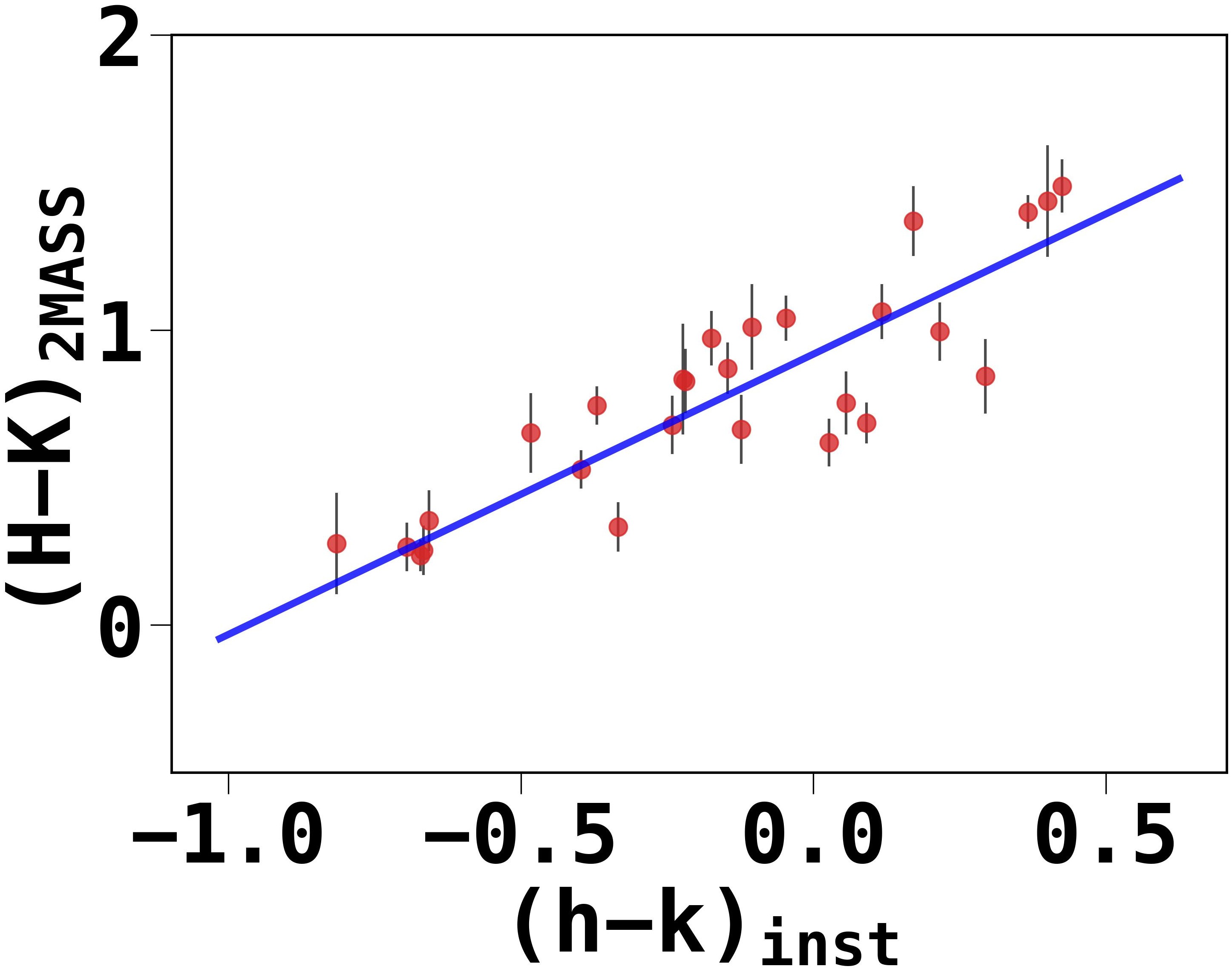}
     \includegraphics[width=4.1cm]{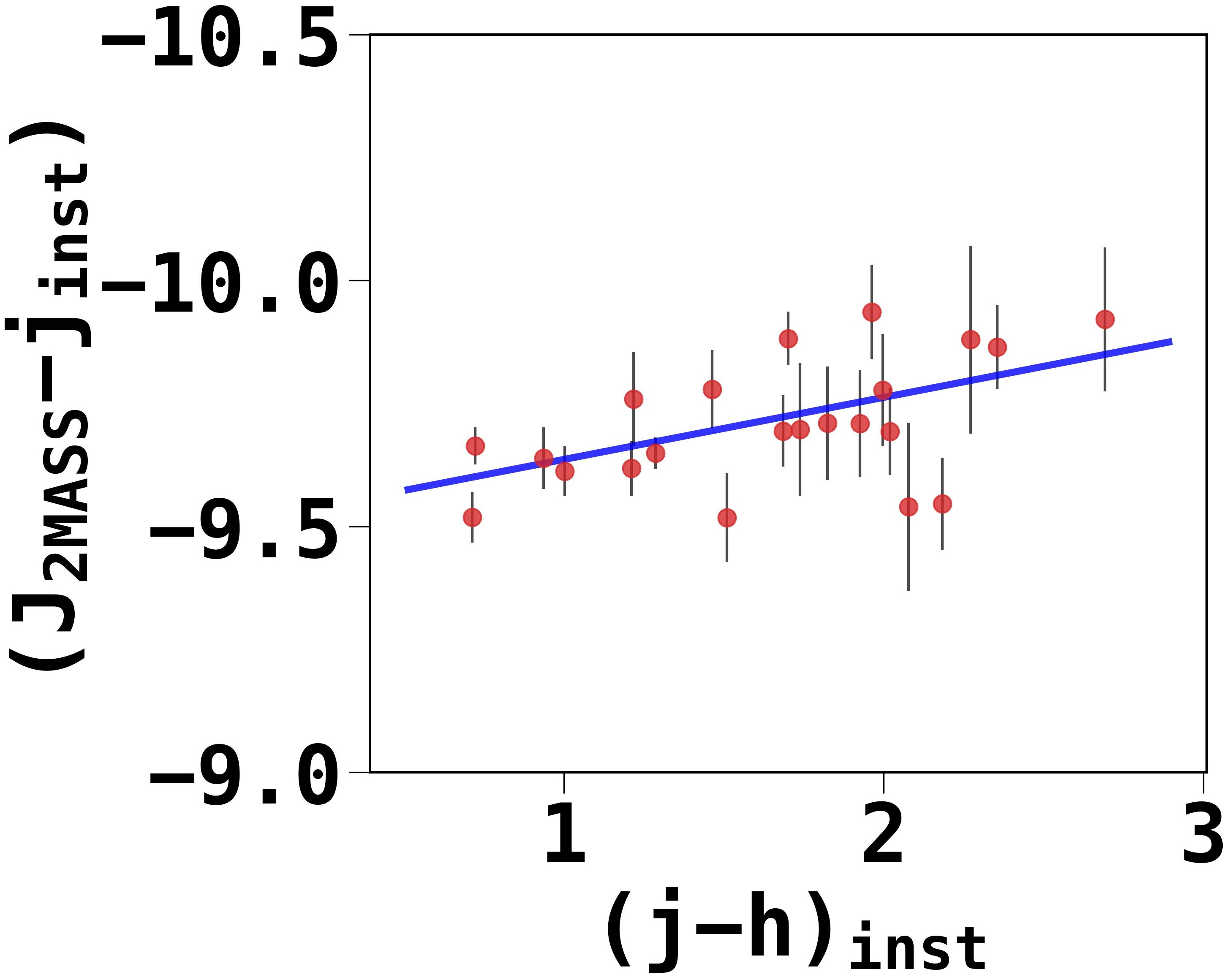}    
 \caption{Color--color plots of 2MASS magnitudes versus TANSPEC instrumental magnitudes in the $J$, $H$, and $K_s$ bands.}
    \label{fig_tans}
\end{figure}


The NIR photometric observations in the $J$ (1.250 $\mu$m), $H$ (1.635 $\mu$m), and $K_s$ (2.150 $\mu$m) bands were carried out on 2022 November 27 and 29 with the 3.6 m DOT telescope \citep{sag19,sag20}, Nainital, India. The observations were taken using the TIFR-ARIES Near Infrared Spectrometer (TANSPEC), mounted at the f/9 Cassegrain focus of the telescope \citep{sha22}. TANSPEC is equipped with a 1k $\times$ 1k HgCdTe imaging array with a pixel scale of 0.245 arcsec, and the image quality is optimized for a 1 arcmin $\times$ 1 arcmin field of view (FOV).

With TANPSEC, we observed the cluster in four pointings, covering $\sim$2 arcmin $\times$ 2 arcmin FOV around the central area of the hub. For each pointing, we employed the seven-point dithered pattern in the $J$, $H$, and $K_s$ bands. In each dithered position, we took 8 frames, with an exposure of 20 seconds per frame. The total integration time of the observation per pointing was about 19 minutes in the $J$, $H$, and $K_s$ bands. 

The standard processing tasks of dark correction, flat-fielding, sky subtraction, and bad-pixel masking were performed. For astrometry, we used the WCS tools and SExtractor\footnote{\href{https://www.astromatic.net/software/sextractor/} {https://www.astromatic.net/software/sextractor/}}, and finally obtained the calibrated, stacked, and mosaicked science images in three bands \cite[for details see][]{ojha04,nei15,shar23}. The FWHM values of the images were in the range of 0.8$-$1.0 arcsec. 

The photometry was done using the packages available in IRAF \citep{iraf_1986, iraf_1993}. Using the DAOFIND task of IRAF, we obtained the list of point sources in the $K_s$-band with signal 5$\sigma$ above the background. 
We performed point spread function photometry of the sources using the ALLSTAR routine of IRAF. For absolute photometric calibration, we used moderately bright and relatively isolated sources from the 2MASS point sources catalog \citep{Skrut_2006} with the quality flag `AAA' and photometric error less than 0.1 mag. 
We obtained the following transformation equations between 2MASS and TANSPEC for the selected sources, which are illustrated in Figure \ref{fig_tans}.

\begin{equation}
    (J-H) = (0.894 \pm 0.069) \times (j-h) - 0.238 \pm 0.131
\end{equation}

\begin{equation}
    (H-K_s) = (0.950 \pm 0.097) \times (h-k_s) + 0.918 \pm 0.039
\end{equation}

\begin{equation}
    (J-K_s) = (0.907 \pm 0.036) \times (j-k_s) + 0.696 \pm 0.056
\end{equation}

\begin{equation}
    (J-j) = (-0.126 \pm 0.040) \times (j-h) -9.510 \pm 0.055
\end{equation}

In the above equations, $J$, $H$, and $K_s$ are the standard magnitudes of
the stars taken from 2MASS, whereas $j$, $h$, and $k_s$ are the instrumental magnitudes from TANSPEC
observations. We applied these transformation equations to all the detected sources in the target field. For sources detected in a single band, we simply applied constant shifts to the instrumental magnitudes to get
the calibrated magnitudes. These constant shifts were determined in each band as the median difference between the instrumental magnitude and the 2MASS magnitudes for the common sources. 
In the present work, we consider sources with errors less than 0.2 mag for the analysis. This 5$\sigma$ sensitivity of the TANSPEC images at the $J$, $H$, and $K_s$ bands are found to be 20.5, 20.1, and 18.6 mag, respectively.

\begin{figure}
    \centering
    \includegraphics[width=8.5cm]{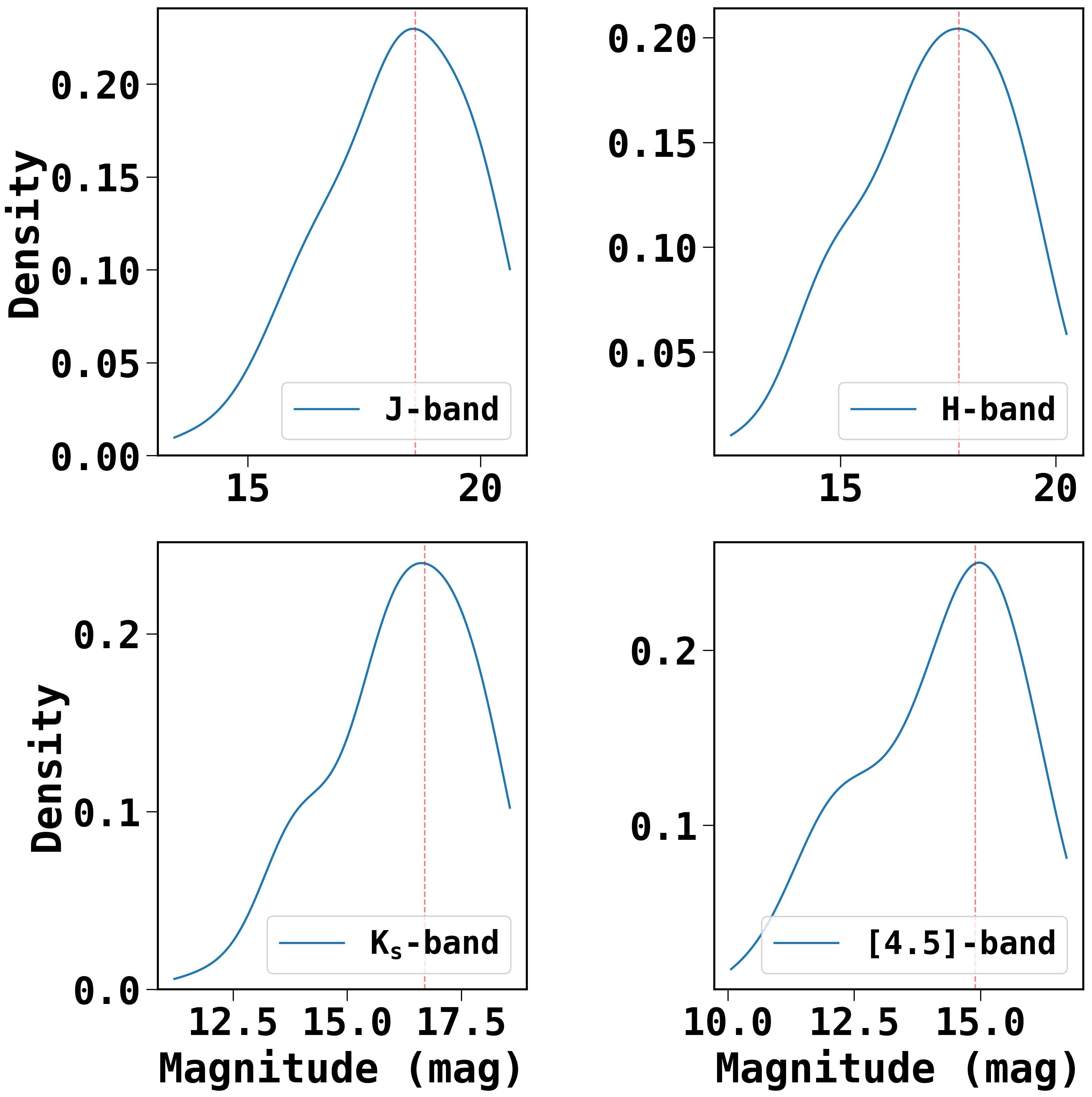}
    \caption{Density plots of photometric data of the cluster region in the $J$, $H$, and $K_s$ bands from the TANSPEC and 4.5\mum band from $\it{Spitzer}$. The dashed lines in all the panels show the completeness limiting magnitudes of 18.6 mag, 18 mag, 17.7 mag, and 14.9 mag in the $J$, $H$, $K_s$, and $[4.5]$\mum bands, respectively.}
    \label{fig_comp}
\end{figure}

\begin{figure*}
    \centering
    \includegraphics[width=18cm]{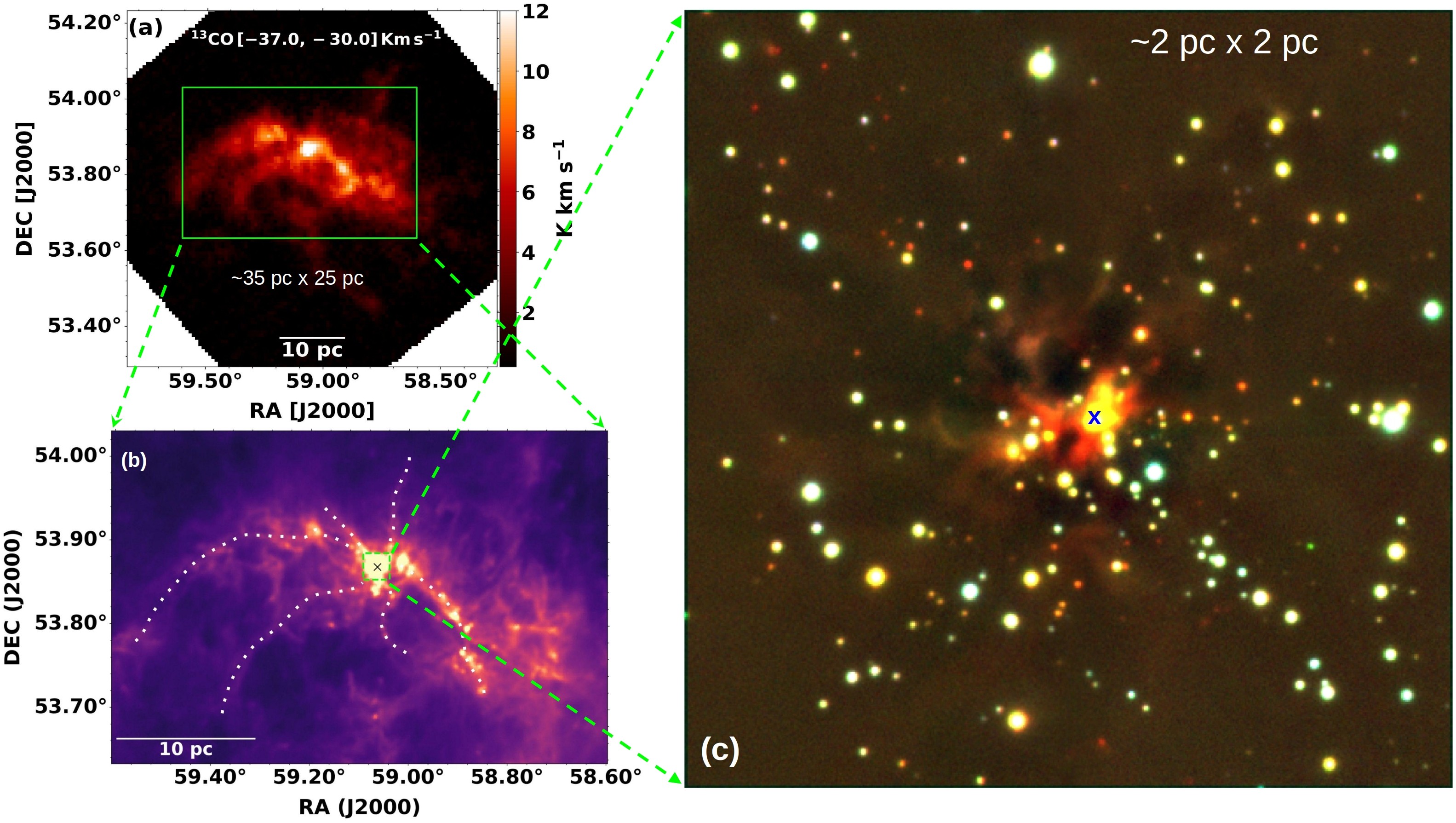}
    \caption{(a) \thco molecular gas distribution of \cloud. The green solid box here shows the inner cloud region zoomed-in in panel-b. (b) $\it{Herschel}$ 250 $\mu$m image of the inner cloud region of size $\sim$35 pc $\times$ 25 pc (marked by a green solid box in panel-a), along with small-scale filamentary structures adopted from \cite{raw23}. The green dashed box shows the hub region of size $\sim$2 pc $\times$ 2 pc, which is observed with TANSPEC. (c) NIR color-composite image (red: $K_s$-band; green: $H$-band, and blue: $J$-band) of \cluster~as seen by TANSPEC. The location of the massive YSO (see text) is shown by a cross symbol.}
    \label{fig_img}
\end{figure*}

\subsection{Galactic Population Synthesis Simulation Data}
\label{sec_pop}
In order to assess the likely contamination of the field population to the cluster population along the line of sight, we obtained the Galactic population in the direction of the cluster using the Besançon population synthesis model\footnote{\href{https://model.obs-besancon.fr/}{https://model.obs-besancon.fr/}} \citep{Robin_2004}. 
To obtain the model population, we simulated the Besançon models for an area equivalent to the observed area of the cluster by adopting the 2MASS photometric system and utilizing the atmosphere models grid of \cite{Allard_2010}. In the simulations, we constrained the photometric errors in 2MASS bands by taking the error as an exponentially increasing function of magnitude, as found in our TANSPEC observations for the cluster. The values of the error function parameters fed to the simulations are obtained by fitting the exponential function over the TANSPEC data. 
For line-of-sight extinction, we used the commonly adopted Galactic extinction value of 1.2 $\rm{mag\,kpc^{-1}}$ \citep{Gont_2012}. 
The Besançon model output data contain the distance, visual extinction, $J$, $H$, and $K_s$ magnitudes, and the spectral type of each synthetic star. We have used this modeled field population to remove the likely contamination present along the line of sight of the cluster (discussed in Sect. \ref{field_pop}). 

\subsection{Completeness of the Photometric Data}\label{phot_comp}
In order to access the overall completeness limits of the photometric catalogs, we use the histogram turnover method \citep[e.g.][]{oh13,sam15, jose17, dam21}. 
In this approach, the magnitude at which the
histogram deviates from the linear distribution is, in general, considered as 90\% complete.
Figure \ref{fig_comp} shows the kernel density estimation (KDE) histograms of the sources detected in various bands. The KDE distribution was done using a multivariate normal kernel,
with isotropic bandwidth $=$ 1.2 mag. This value was chosen as we find that it is a good compromise between over- and under-smoothing density fluctuations. 
With this approach, our photometry is likely to complete down to $J$ $\sim$18.6 mag, $H$ $\sim$18 mag,
$K_s$ $\sim$17.7 mag. We also use the $\it{Spitzer}$ 4.5\mum catalog from the GLIMPSE360 survey \citep{Whitney_2008} to access the YSOs of the studied region. The completeness limit of the 4.5\mum catalog is 14.9 mag and is also shown in Figure \ref{fig_comp}. The observations were taken as part of the Spitzer Warm Mission Exploration Science program and performed using the two short-wavelength IRAC bands at 3.6 and 4.5\mum.   

 


\section{Results and Discussion}
\label{analysis}
\subsection{Overview of the \cloud~in CO}

\begin{figure*}
    \centering{
    \includegraphics[height=6.5cm]{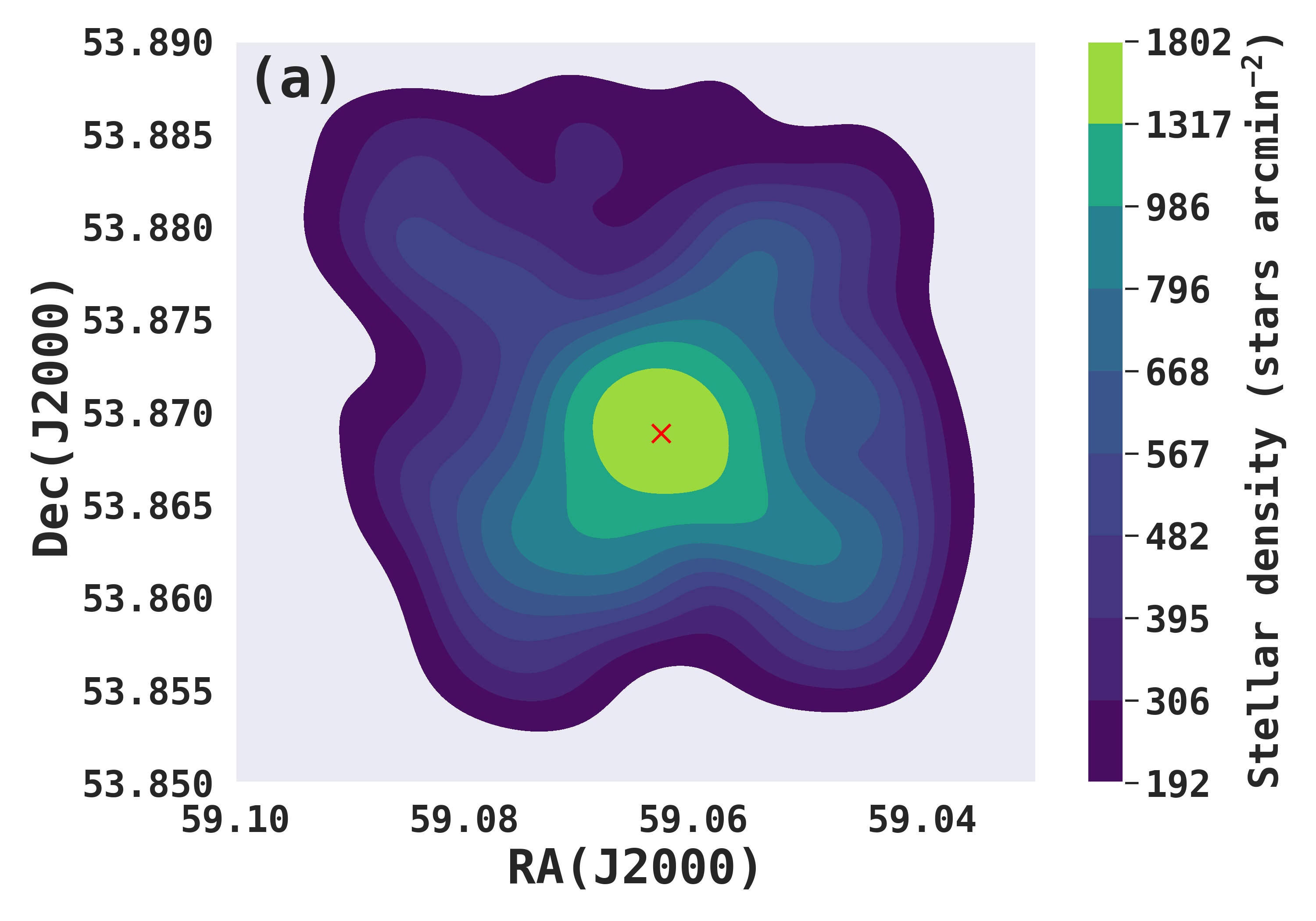}
     \includegraphics[height=6.5cm]{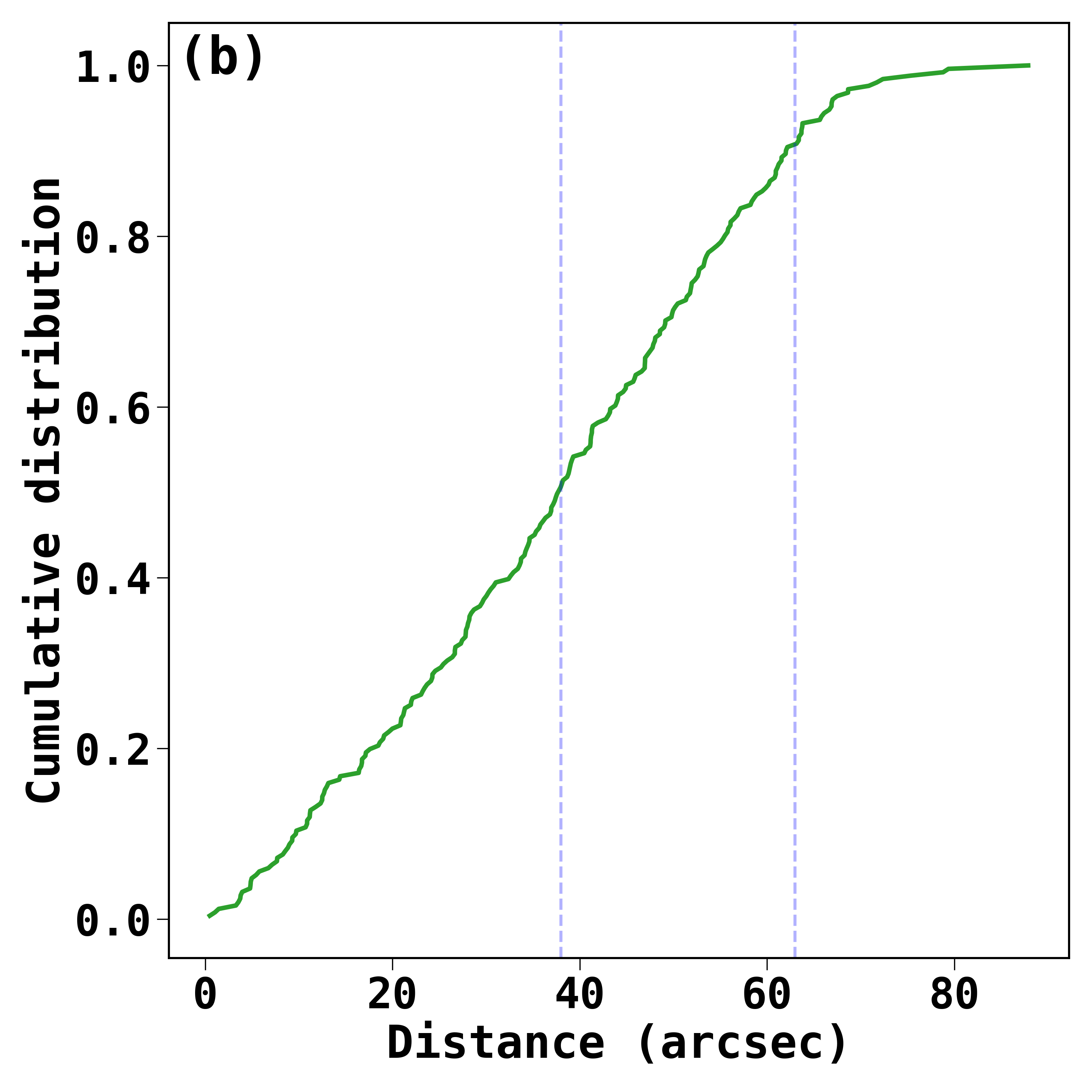}}    
    \caption{
    (a) Smoothened 2D density map of the sources, shown from the peak density up to the 10\% level of stellar density. (b) Cumulative distribution of all the sources as a function of distance from the cluster center (marked by a cross in panel-a). The dashed lines show the distances from  the cluster center 
    within  which 50\% and 90\% of the sources are lying.}
    \label{fig_dist}
\end{figure*}

Figure \ref{fig_img}(a) shows the distribution of the \thco integrated emissions of \cloud~as observed with the Purple Mountain Observatory (PMO) 13.7 m telescope (beam $\sim$52\arcsec). Details of the CO observations and the intensity map can be found in \cite{raw24a}. 
The \thco emission represents the relatively dense inner area of the cloud with an effective radius of $\sim$17 pc (at $d$ $\sim$3.4 kpc). From Figure \ref{fig_img}(a), it can be seen that there is a bright spot at the heart of the cloud (i.e., near the geometric 
center). Based on \eico observations, \cite{raw24a} observed that this bright spot corresponds to the location of the most massive clump of the cloud, onto which several large-scale filaments are funneling cold gaseous matter. The small-scale filamentary structures attached to the center of the hub, as found in \cite{raw23}, are
shown in Figure \ref{fig_img}(b), mimicking the hub-filament system morphology as found in other star-forming regions \citep[e.g.][]{mye09, kum22}.

\subsection{Stellar Content and Cluster Properties}

Figure \ref{fig_img}(c) shows the NIR image of the hub as seen in the TANSPEC bands. We find that the clump lacks point sources in the optical bands (e.g., in Digitized Sky Survey's images), while in NIR images, clustering of point sources along with infrared nebulosity can be seen. 
Figure \ref{fig_dist}(a) shows the 2D density map of the point sources in the studied area. 
The stellar density is shown by shaded colors from the peak density to the 10\% level. Figure \ref{fig_dist}(b) shows the cumulative distribution of stars from the center of the density distribution, and as can be seen, stars are spread within $\sim$90\arcs~radius from the cluster center, with 50\% lying within 38\arcs~and 90\% within
63\arcs. Beyond 63\arcs, the distribution deviates from the linear shape and becomes flatter, implying that the improvement in the cluster density is insignificant beyond 63\arcs. We thus considered 63\arcs~as the conservative radius of the cluster, which is around 1 pc at the distance of \cloud~(i.e., $\sim$3.4 kpc). 
  
The cluster is embedded in a cloud of visual extinction as high as 20$-$30 mag \citep{raw23}, therefore, the contamination due to background stars to the cluster members is expected to be low. 
Below, we estimate the likely contamination of the field stars and derive properties of the cluster such as the extinction, $K_s$-band luminosity function, age, mass function, and star-formation efficiency and rate. 

\subsubsection{Extinction}
\label{extin}

\begin{figure}
\centering{
\includegraphics[width=8cm]{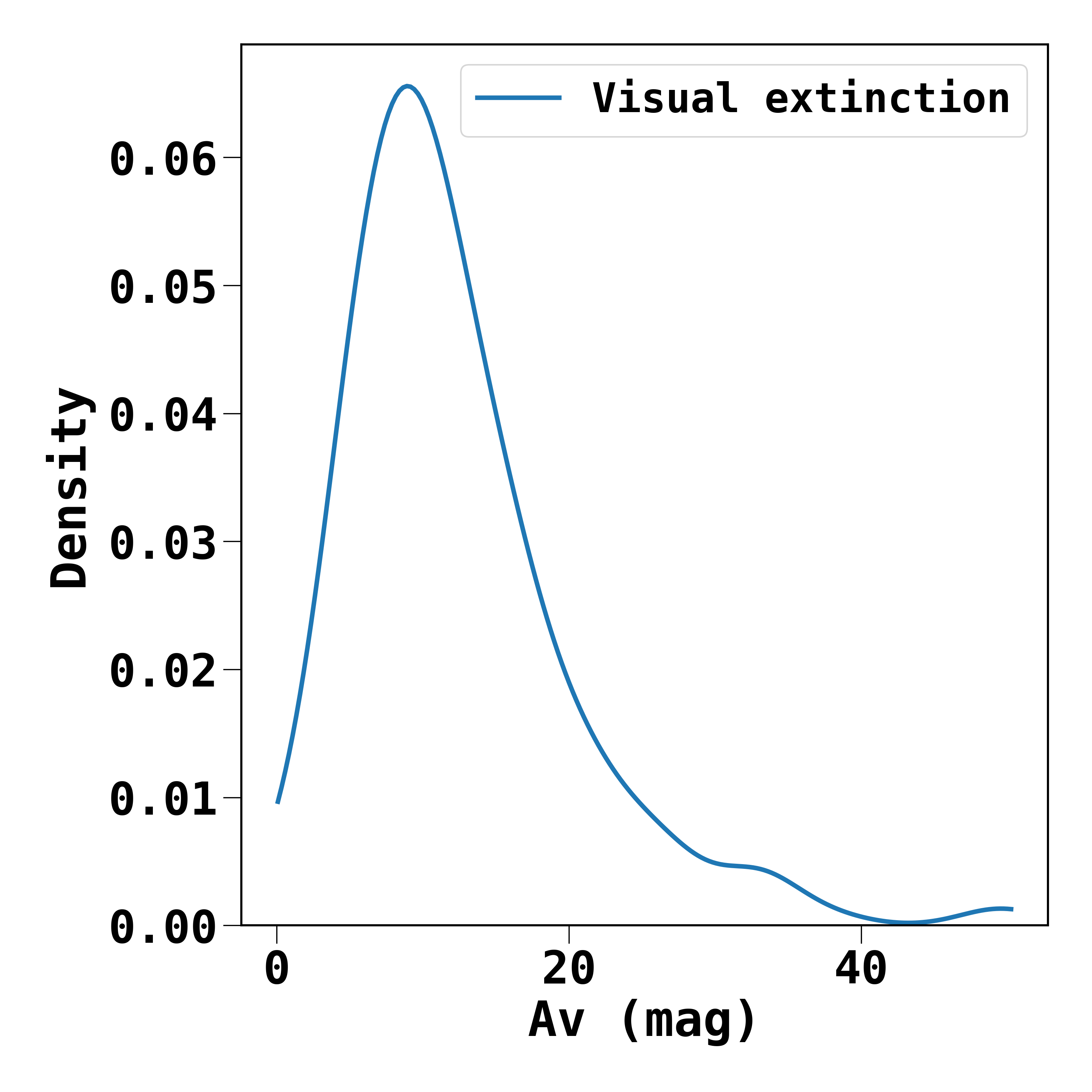}}
\caption{Density plot of all the observed sources in the cluster as a function of their visual extinction ($A_{\rm{V}}$; see Section \ref{extin}). }
\label{fig_av}
\end{figure}

Extinction plays an important role in deriving cluster properties. The line-of-sight extinction of an individual star can be 
directly determined from knowledge of its color excess 
and the extinction law. We estimated the extinction of all the stars observed toward the cluster direction within its radius ($\sim$1\arcm) using the relation 
\begin{equation}
A_V = c \times [(i - j)- (i-j)_0],
\label{eq:ext}
\end{equation}
where $(i-j)$ is the apparent and $(i-j)_0$ is the intrinsic colors of the point sources in $i^{th}$ and $j^{th}$ filter. Here, $c$ is the constant based on the extinction law of \citet{rie85}, which is 9.34 and 15.98 for the $J-H$ and $H-K_s$ color excess, respectively. In general, the $H-K_s$  colors are preferred for deriving extinction of star-forming regions \citep[e.g.][]{gut09} because, in such environments, the majority of the sources can be highly embedded in the dust to be detected in the $J$-band.   
However, excess emission due to circumstellar disks from young sources (see Appendix \ref{sec_cc} for the discussion on the identification of IR-excess sources) can significantly impact the $H-K_s$  colors, causing them to appear redder than their intrinsic photospheric colors, resulting in higher \av values. This can be significant in young clusters, where a significant fraction of the stars can have a circumstellar disk. We thus used both $J-H$ and $H-K_s$ colors to determine the \av of the observed sources by assuming that the majority of the sources within the cluster boundary are cluster members. We then made a combined \av catalog, where priority was given to the \av values obtained from the $J-H$ colors of the common sources, else \av values from the $H-K_s$  colors were considered. For estimating $A\rm{_V}$, we use the median intrinsic $J-H$ and $H-K_s$  colors of the GKM dwarfs\footnote{\href{https://www.pas.rochester.edu/~emamajek/EEM_dwarf_UBVIJHK_colors_Teff.txt}{Stellar Color/Teff Table}}  as 0.39 mag and 0.14 mag from \cite{pec13} as the typical intrinsic color of the point sources. 
The \av distribution, with a peak around 11 mag, is shown in Figure \ref{fig_av}. 
Although some sources show high \av value, we find that \av $<$ 22 mag encompasses the majority ($\sim$90\%) of the sources. For sources within \av $<$ 22 mag, we find that the resultant median visual extinction is 11 $\pm$ 4 mag, whose corresponding extinction at $K_s$-band ($A_{\rm{K}}$) is 
$\sim$1.23 $\pm$ 0.40 mag, following the extinction law ($A\rm{_K} = 0.112 \times A_{\rm{V}}$) of \citet{rie85}.  

We find that the median \av for the common sources detected in the $H$ and $K_s$ bands is higher by $\sim$1.3 mag compared to the sources detected in the $J$ and $H$ bands. Assuming that the inner-disk emission dominates in the $K_s$-band and has minimal contribution in the $H$ band, we attributed this excess extinction could be due to the emission from
the circumstellar disks of the disk-bearing cluster members. This excess extinction is equivalent to $\Delta A\rm{_K} $ $\sim$0.15 mag, in the $K_s$-band. We discuss more on this excess extinction in Section \ref{KLF_age}. 

We acknowledge the fact that although the individual \av values are derived using the extinction law of \cite{rie85}, they only accurately reflect the true visual extinctions as long as the assumed reddening law is appropriate for this cloud. The extinction law given by \cite{rie85} has a negative power-law dependency on the wavelength with a power-law slope, alpha $\sim$1.6.
Grain growth in cold clouds can alter the extinction law. Some studies show higher alpha values, 1.6$-$2.6, with a median around 1.9 in molecular clouds \citep[see][and references therein]{Wang_2014, Maiz_2024}.  
If we use the extinction law corresponding to the slope of 1.9 \citep{messineo_2005}, we find that the median \av of the cluster changes by only $\sim$0.3 mag (or \ak = 0.03 mag).

In the preceding paragraphs, we determined the median \av of the cluster by assuming that all the observed sources within the cluster boundary are cluster members. However, if the field population is significant toward the cluster direction, it may affect the true median extinction value of the cluster. To further validate the robustness of the derived extinction value,
we use the NIR-excess-emission sources, identified 
using the $JHK_s$ and $HK_s[4.5]$ color-color (CC) diagrams
(discussed in Appendix \ref{sec_cc}) for deriving the median extinction of the cluster. Using only these excess sources (i.e., disk-bearing cluster members) and following the same approach illustrated in the previous paragraphs, we find that the median visual extinction turns out to be around 11 mag, in agreement with the earlier estimation. We also find that using an extinction law of alpha $\sim$1.9, the disk fraction remains the same within the error. Since the use of a higher alpha value is not changing the results in a major way, we continue further analysis with the extinction law of \cite{rie85} for the general interstellar medium.


\subsubsection{Likely Field Population}
\label{field_pop}

\begin{figure}
    \centering
    \includegraphics[width=8cm]{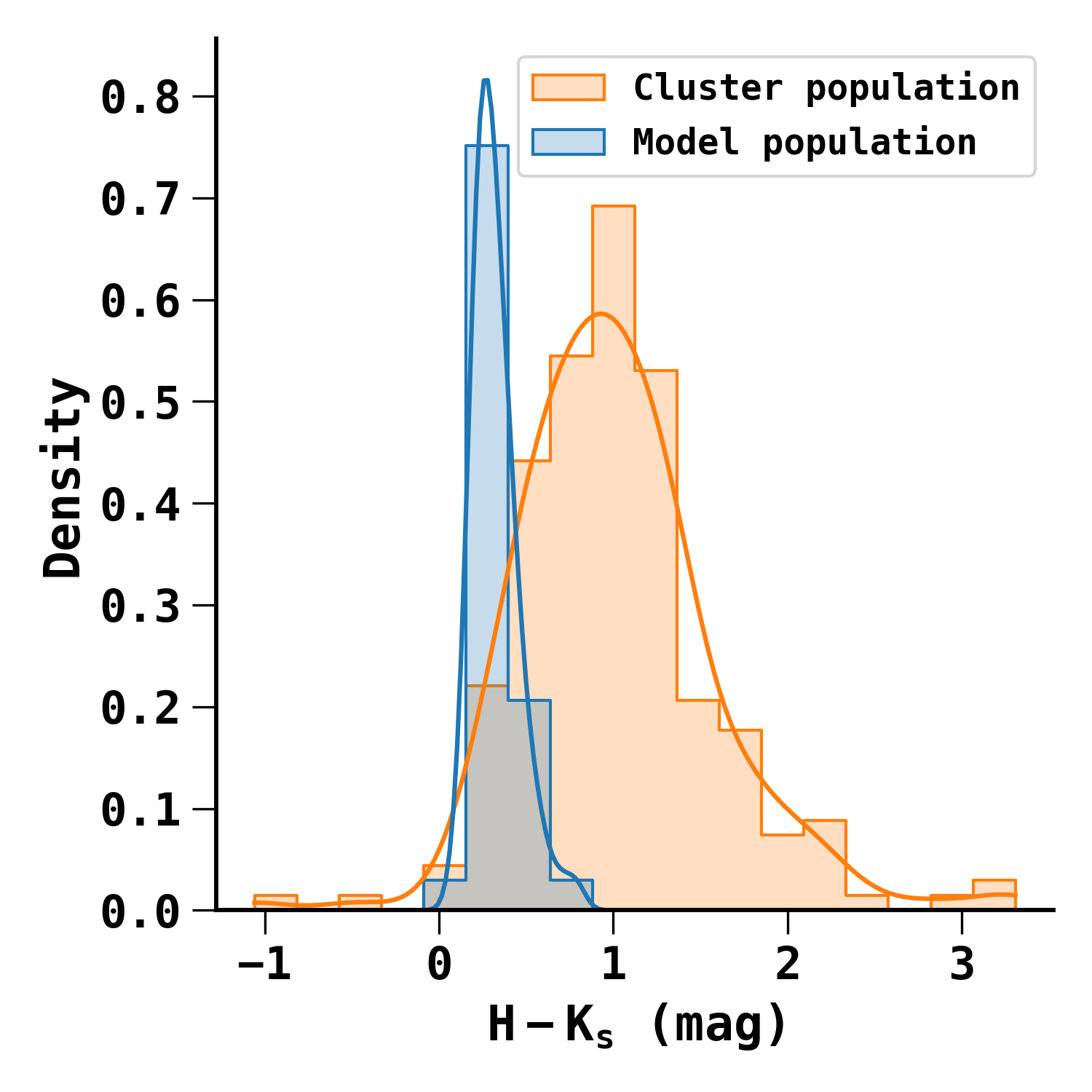}
    \caption{Distribution of all the sources observed toward the cluster and the Besançon model generated sources, shown as a function of their ($H-K_s$) colors. The blue and orange curves show the corresponding density plots of the cluster and model population, respectively.}
    \label{fig_hk_color}
\end{figure}

\begin{figure}
    \centering
    \includegraphics[width=8cm]{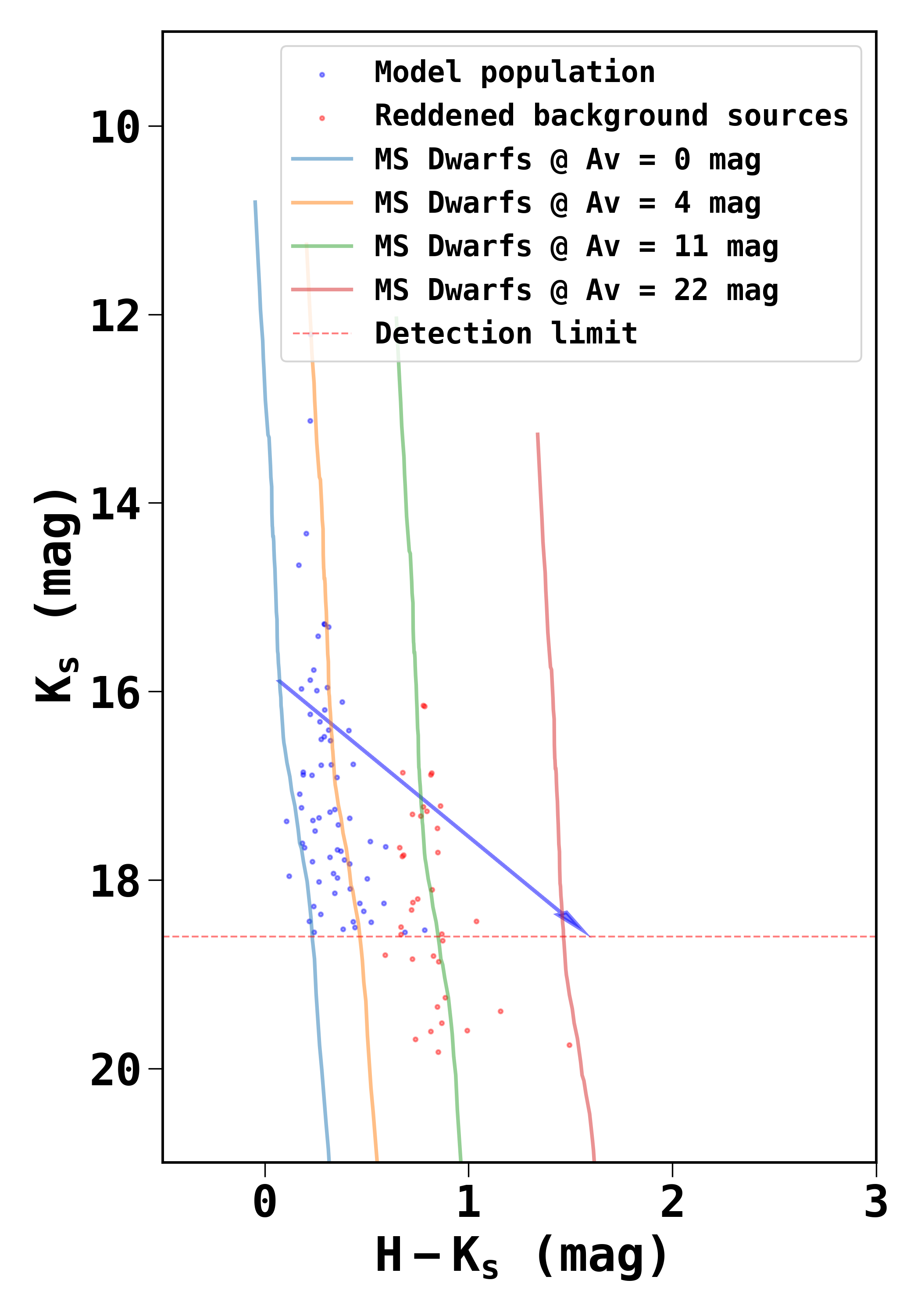}
    \caption{$K_s$ versus $H-K_s$ diagram of the model population. The red dots show the background sources ($d$ $>$ 3.4 kpc) reddened by \av = 11 mag (average extinction toward the cluster). The colored curves show the main-sequence dwarfs locus reddened by \av = 0, 4, 11, and 22 mag. The blue arrow shows the reddening vector drawn from the 1 \Ms~limit. }
    \label{fig_hk_color2}
\end{figure}

Figure \ref{fig_hk_color} shows the $H-K_s$ color distribution of the field sources obtained from the population synthesis model (see Section \ref{sec_pop}), as well as of the total observed sources in the cluster direction. As can be seen, the field population shows a narrow $H-K_s$ color distribution peaking at 0.3 mag (i.e., corresponds to \av $\sim$3 mag, using equation \ref{eq:ext}), with the majority lying below 0.4 mag (i.e., corresponds to \av $\sim$4 mag) while the $H-K_s$ color of the cluster field shows a wide distribution having a peak around 1 mag, with the majority lying below 2.5 mag. It implies that the majority of the sources with $H-K_s$ color below 0.4$-$0.5 mag are likely the field population along the direction of the cluster. From the population
synthesis model, we find that the majority of the model population is located
at a distance of less than 3.4 kpc and, thus, is likely the foreground population
in the direction of the cluster. There may be some background field stars within our cluster sample, but we assume their contribution to be small, given the sensitivity of our observations and the high
column of matter present in the clump. 
To illustrate
this, in Figure \ref{fig_hk_color2}, we show the $K_s$ versus $H-K_s$ diagram
of the population synthesis field sources along with the main-sequence locus reddened by \av = 0, 4, 11, and
22 mag. As can be seen,  most of the relatively bright stars (e.g., $K_s$ $<$ 15.5 mag) and the majority of the faint stars are located within the \av = 4 mag locus, suggesting 4 mag is likely the foreground extinction in the direction of the cluster. The red stars represent the background sources (i.e., sources with $d$ $>$ 3.4 kpc) reddened by \av = 11 mag and \av = 22~ mag. As can be seen, if background sources are located behind \av = 22 mag, most of them would be beyond our sensitivity limit of the $K_s$-band, while only a few sources would contaminate our cluster sample if they are located behind the cluster and lie in the range of \av $\sim$11$-$22 mag.  

If we assume that most of the sources within the cluster radius 
with color $H-K_s$ $>$ 0.5 mag are likely the cluster members,  we expect the percentage of background contamination in our sample to be around  $\sim$10\%.  We discuss more on this point further in Section \ref{sec_mass_ext}. This contamination fraction 
would be further less if we consider sources above 0.5$-$1 \Ms. For example, the blue arrow in Figure \ref{fig_hk_color2} shows the reddening vector from the base of 1 \Ms~dwarf locus, which reveals that the background contamination above 1 \Ms~is negligible.


\subsubsection{$K_s$-band Luminosity Function and Likely Age}
\label{KLF_age}
\begin{figure*}
    \centering{
     \includegraphics[width=8 cm]{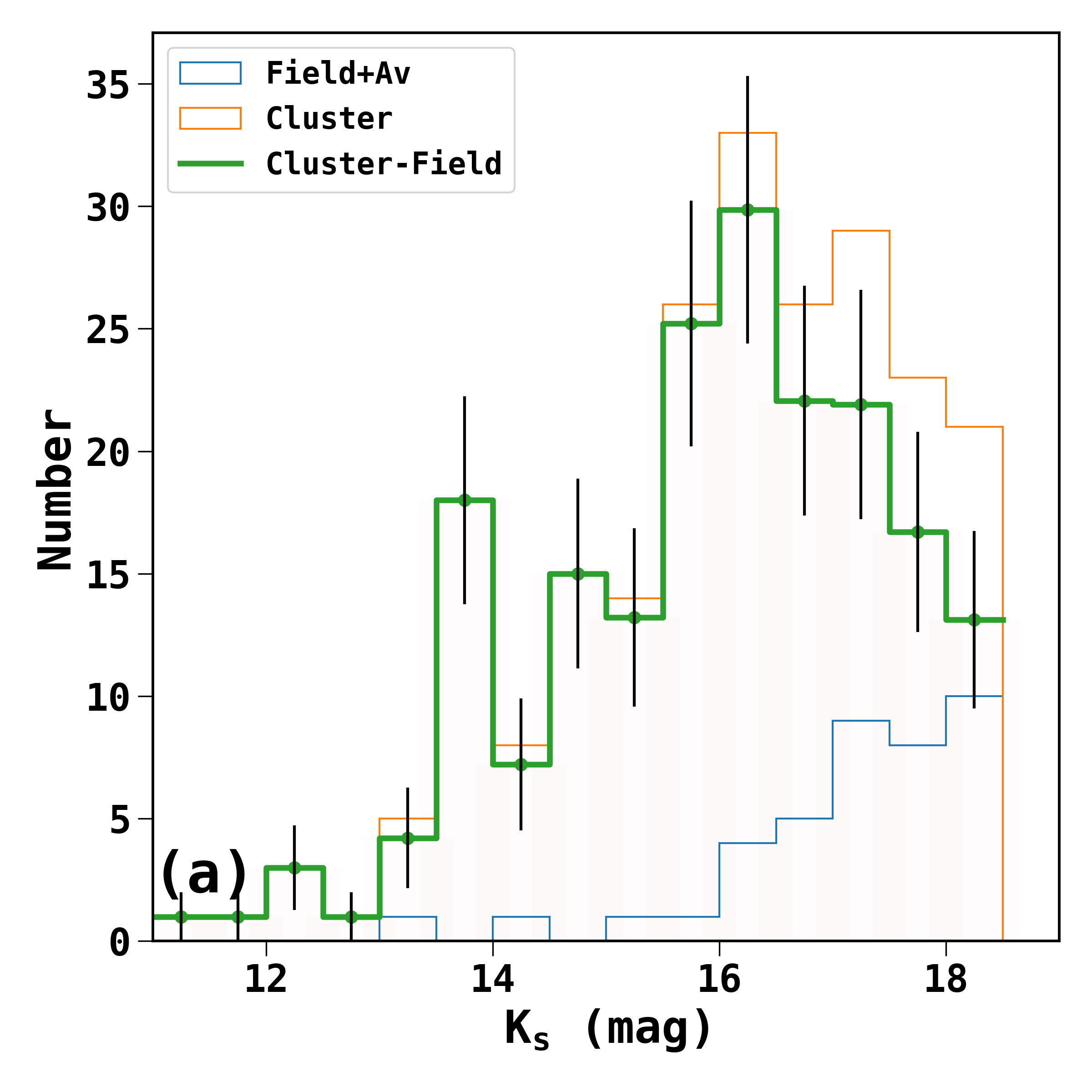}
    \includegraphics[width=8.5 cm]{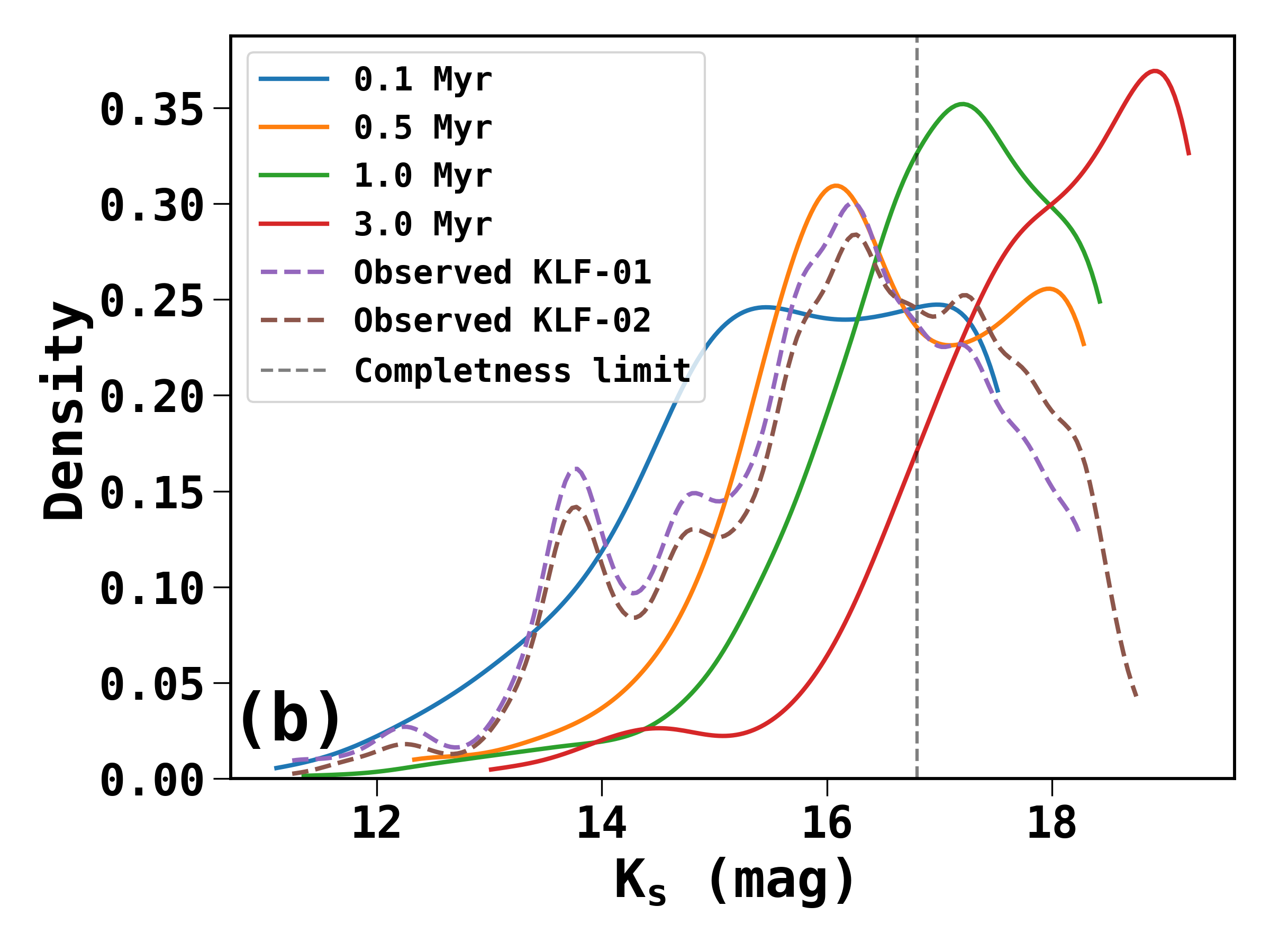}}
    \caption{(a) $K_s$-band luminosity function of the cluster, reddened model control field, and control field subtracted cluster shown by orange, blue, and green histograms, respectively. The error bars represent the Poisson error. (b) $K_s$-band density plots of synthetic clusters of age 0.1, 0.5, 1.0, and 3.0 Myr, shown by solid curves. The dashed curves show the $K_s$-band density plots of the reddened control field subtracted cluster.}
    \label{fig_klf}
\end{figure*}

\begin{figure}
    \centering
    \includegraphics[width=8.5cm]{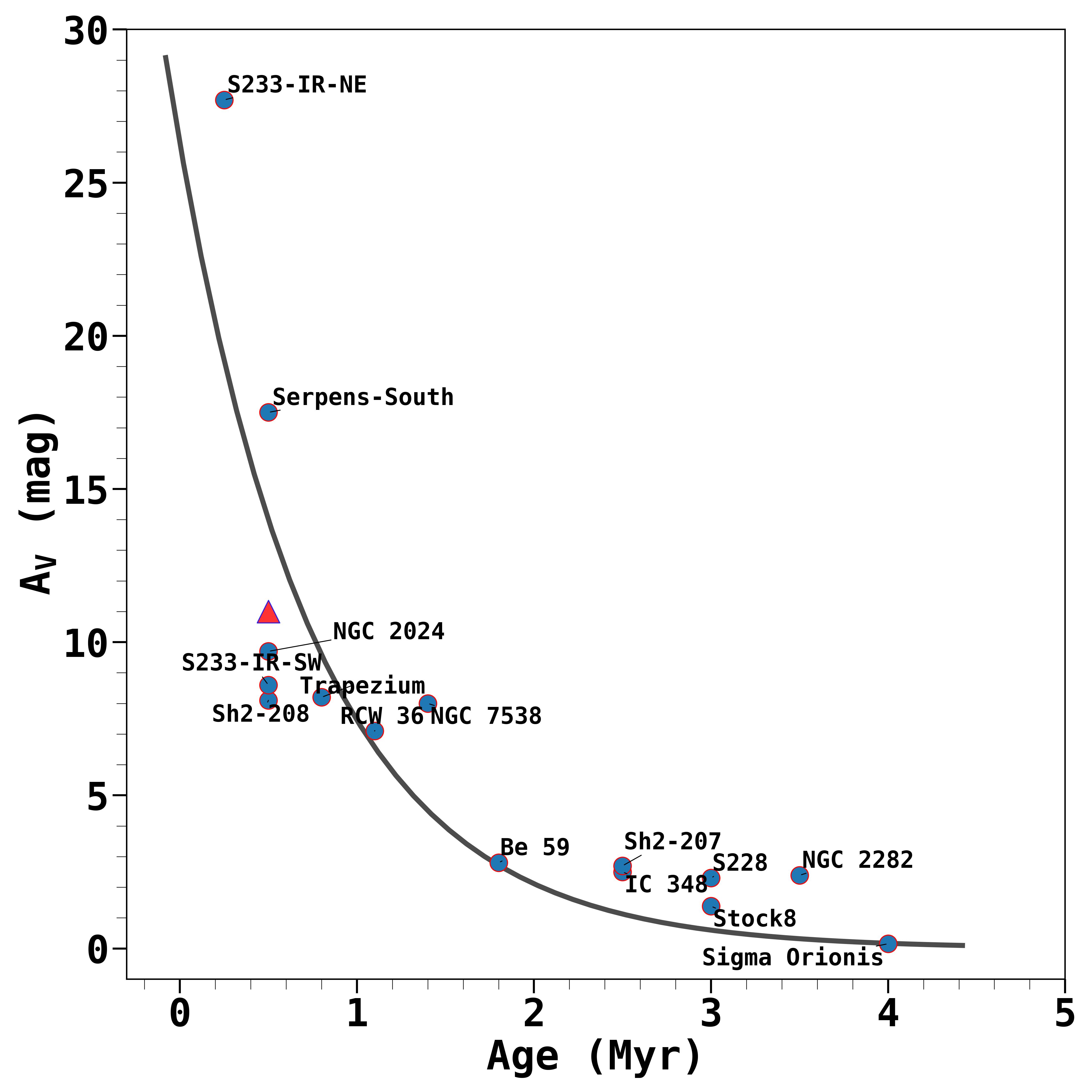}
    \caption{Visual extinction versus age plot of different nearby clusters (d $\lesssim$ 4 kpc), given in table \ref{tab:sfr}. The black curve shows the best-fit exponential function (see text for details) with fitted parameters, $a$ $\approx$ 26.30 $\pm$ 3.33 and $b$ $\approx$ 1.26 $\pm$ 0.05. The red triangle shows the position of \cluster. Here, the extinction values are corrected for foreground extinction, as found in the literature.}    
    \label{fig_av_age}
\end{figure}

The $K_s$-band luminosity functions (KLFs) of different ages are known to have different peak magnitudes and slopes. Thus, a comparison of the observed KLF with the model KLFs can constrain the age of a cluster \citep{lada95,meg96, mue00,ojha04, ojha11, jose12, Brajesh_2014}. KLF is expressed by the following
equation:
\begin{equation}
\frac{dN}{dm_k} = \frac{dN}{dM_*} \times \frac{dM_*}{dm_k}, 
\end{equation}
where $m_k$ is the $K_s$-band luminosity and $M_*$ is the stellar mass \citep[e.g.,][]{lada03}. In the equation, the left-hand term represents the number of stars for a given $K_s$-band magnitude bin, while the first term on the right-hand side is the underlying stellar mass function, and the second term is the mass-luminosity ($M-L$) relation.
To derive the KLF of the cluster, one first needs to correct for field contamination. To do so, we used the model star counts predicted by the Besançon model discussed in Section \ref{sec_pop}. The advantage of using the Besançon model is that the background stars ($d$ $>$ 3.4 kpc) can be separated from the foreground stars ($d$ $<$ 3.4 kpc).
 While all the stars in the field suffer a general interstellar extinction, only the background stars suffer an additional extinction due to the molecular cloud. Besançon model gives extinction of individual stars ($A_{\rm{Vi}}$) along the line of sight. The median extinction of
 the cluster is \av=11 mag.  We, therefore, reddened the background stars by applying an extra extinction of $\Delta A_{\rm{Vi}}$ (= \av $-$ $A_{\rm{Vi}}$) to put them behind a cloud of visual extinction 11 mag. We then combined both the foreground and reddened background stars to make a whole set of contaminating field stars. Then, we correct cluster counts by subtracting the contaminating field star counts from the cluster star counts. The field-corrected cluster KLF, which we called ``KLF-01,'' is shown in  Figure \ref{fig_klf}(a). The figure also shows the KLF of the reddened field and cluster before the field decontamination population. We also made another field-corrected KLF, which we called ``KLF-02.'' The KLF-02 is obtained by simply reddening all the model sources by \av = 8 mag (\av = 11 $-$ 3), thereby bringing both the cluster and field sources to the same median \av value of 11 mag, and then subtracting the field counts from the cluster counts. The 3 mag is the average extinction of the field sources based on their average $H-K_s$ colors (see Section \ref{field_pop}). The second method is also often adopted in the literature when distance information of the 
 field sources is not available \citep[e.g.][]{jose11}.

We then generate synthetic clusters for the age range between 0.1 and 3 Myr at an interval of 0.5 Myr, using the Stellar Population Interface for Stellar Evolution and Atmospheres (SPISEA) Python code \citep{Hosek_2020}. We choose the following procedure in the SPISEA code: (1) assume that the distribution of stars in the cluster follows the \cite{kro01} mass function, (2) use the $M-L$ relation for the aforementioned ages from the MIST isochrone models of solar metallicity \citep{choi_2016}, and (3) adopted the Rieke-Lebofsky extinction laws \citep{rie85} and 2MASS filter passbands \citep[for more details, see][]{Hosek_2020}. Next, we convert the absolute $K_s$-band magnitude of the cluster stars to the apparent magnitude using the distance modulus and average extinction of \ak $\sim$1.2 mag (\av $\sim$11 mag) of the cluster. Then, we constructed the KLFs using apparent $K_s$-band magnitudes and subsequently smoothed the KLFs with a Gaussian KDE bandwidth of $K_s$ = 1 mag, to account for various uncertainties present in the observed cluster. These uncertainties include the uncertainty of $\sim$0.4 mag in the mean extinction of the cluster and a likely uncertainty of $\sim$0.15 mag due to excess extinction (see Section \ref{extin}), in the $K_s$-band. The last step
is done to compare model KLFs with the KLF of the observed cluster. Since SPISEA randomly generates sources for a cluster, we ran the simulation 200 times for each synthetic cluster age. Then we obtained the median KLF for each synthetic cluster. Figure \ref{fig_klf}(b) shows the model KLFs of different ages along with the observed
field-star-subtracted cluster KLFs (KLF-01 and KLF-02) derived in two ways discussed above. As
can be seen, barring the bump around $K_s$ $\sim$13.8 mag, the model KLFs with ages in the range of 0.1$-$1 Myr appear to be reasonably matching with the observed KLFs, while the overall shape of the observed KLF is in better agreement with the model KLF of age = 0.5 Myr. Also, from the disk fraction of FSR 655 (discussed in Appendix \ref{disk_frac}), the age of the cluster seems to be less than a Myr. Therefore, $\sim$0.5 Myr appears
to be a reasonable assumption as the age of the cluster.
The reason for the bump in the KLF around  $K_s$ $\sim$13.8 mag is unclear to us, but we believe that given the small area investigated in this work, the origin of the bump is more of a statistical nature.
Wider and deeper observations of the cluster field, as well as a nearby control field, would be able to shed more light on this issue. 

In molecular clouds, gas is either consumed in the star formation processes or dissipated by various feedback effects due to forming stellar members. It has been
found that molecular clouds with an age greater than $\sim$5 Myr are seldom associated with molecular gas \citep{Leisawitz_1989}. So, the lack of molecular gas and dust is a proxy indication of the cloud's evolution. In Figure \ref{fig_av_age}, we show the median visual extinction associated with some of the compact (radius $<$ 3 pc) nearby young clusters ($<$ 4 kpc) of age less than 5 Myr that are associated with a few O-type to early B-type stars. We restrict our sample to the aforementioned type clusters in order to be able to compare with the cluster investigated in this work. 
As one can see from the figure, the visual extinction is decreasing with the age of the cluster, as expected. Seeing the nature of the plot, we fitted the data points with an exponential decay function of the form, \av = $a \times \exp{(-b \tau)}$, where $\tau$ is the age of the cluster. Before fitting, the extinction values are first corrected for foreground extinction, as found in the literature. The best-fit values of $a$ and $b$ are $\sim$26.30 $\pm$ 3.33 and $\sim$1.26 $\pm$ 0.05, respectively. We note that though this oversimplified approach suggests a decrease in the column density exponentially with time, the real scenario might be more complex as it strongly depends 
upon the strength of feedback from the stars present in the clump and the rate of star formation. 
A better sample with nearly similar cluster mass may provide better results; 
nonetheless,  the obtained result provides a proxy way of seeing how the column density might have 
evolved in the clumps that are host to low-to-intermediate-mass clusters, like the one investigated in the present work. As it can be observed from the figure, the median \av of \cluster~is certainly higher than clusters of age older than 2 Myr (e.g., Stock 8, IC 348, and S228) and comparable to the extinction of the clusters in the range 0.5$-$1 Myr (e.g., NGC 2024, Sh 2-208, and  S233-IR-SW). This again points to the fact that the studied cluster is unlikely to be older than a Myr. We also found that the disk fraction of the cluster is compatible with other nearby clusters of younger age less than a Myr (for details, see Appendix \ref{disk_frac}).

\begin{table}

\begin{adjustwidth}{-1.25cm}{}
\begin{threeparttable}
\caption{Parameters of nearby clusters.}

\begin{tabular}{p{0.2cm} p{1.4cm} p{0.6cm} p{0.6cm} p{0.9cm} p{2.5cm}} 

\hline
\hline

No & Name & \av & Age & Distance & References \\ 
\hline
  &        & (mag) & (Myr) & (kpc)  &\\
1 & Stock8 & 2.0 & 3.0 & 2.3 & \cite{jose17,dam21}\\ 
2 & Be 59 & 4.0 & 1.8 & 1.0 & \cite{pan18}\\
3 & S228 & 3.3 & 3.0 & 3.2 & \cite{yad22}\\
4 & IC 348 & 3.5 & 2.5 & 0.32 & \cite{Muench_2007}\\
5 & Trapezium & 9.2 & 0.8 & 0.4 & \cite{mue02}\\
7 & Sh2-208 & 10.1 & 0.5 & 4.0 & \cite{yas16b}\\
8 & Sh2-207 & 2.7 & 2.5 & 4.0 & \cite{yas16a}\\
9 & S233-IR-SW & 9.8 & 0.5 & 1.8 & \cite{yan10}\\
10 & S233-IR-NE & 28.9 & 0.25 & 1.8 & \cite{yan10}\\
11 & NGC 2282 & 4 & 3.5 & 1.65 & \cite{dut15}\\
12 & NGC 7538 & 11 & 1.4 & 2.7 & \cite{sharma17}\\
13 & RCW 36 & 8.1 & 1.1 & 0.7 & \cite{baba04, ell13}\\
14 & NGC 2024 & 10.7 & 0.5 & 0.42 & \cite{lev06}\\
15 & Serpens South & 19.5 & 0.5 & 0.44 & \cite{Jose_2020}\\
16 & Sigma Orionis & 0.155 & 4.0 & 0.4 & \cite{Walter_2008}\\
\hline
\hline

\end{tabular}
\begin{tablenotes}[para]
\setlength{\parindent}{1.1cm}
\hangindent=1.1cm 
\footnotesize{The mean extinction values of the clouds are taken from the quoted references. For the age of the regions, wherever the range was given, we have taken the mid values.}

\end{tablenotes}
\end{threeparttable}
\end{adjustwidth}
\label{tab:sfr}
\end{table}


\subsubsection{Mass-extinction-limited Sample}
\label{sec_mass_ext}
\begin{figure}
\centering{
\includegraphics[width=8.5cm]{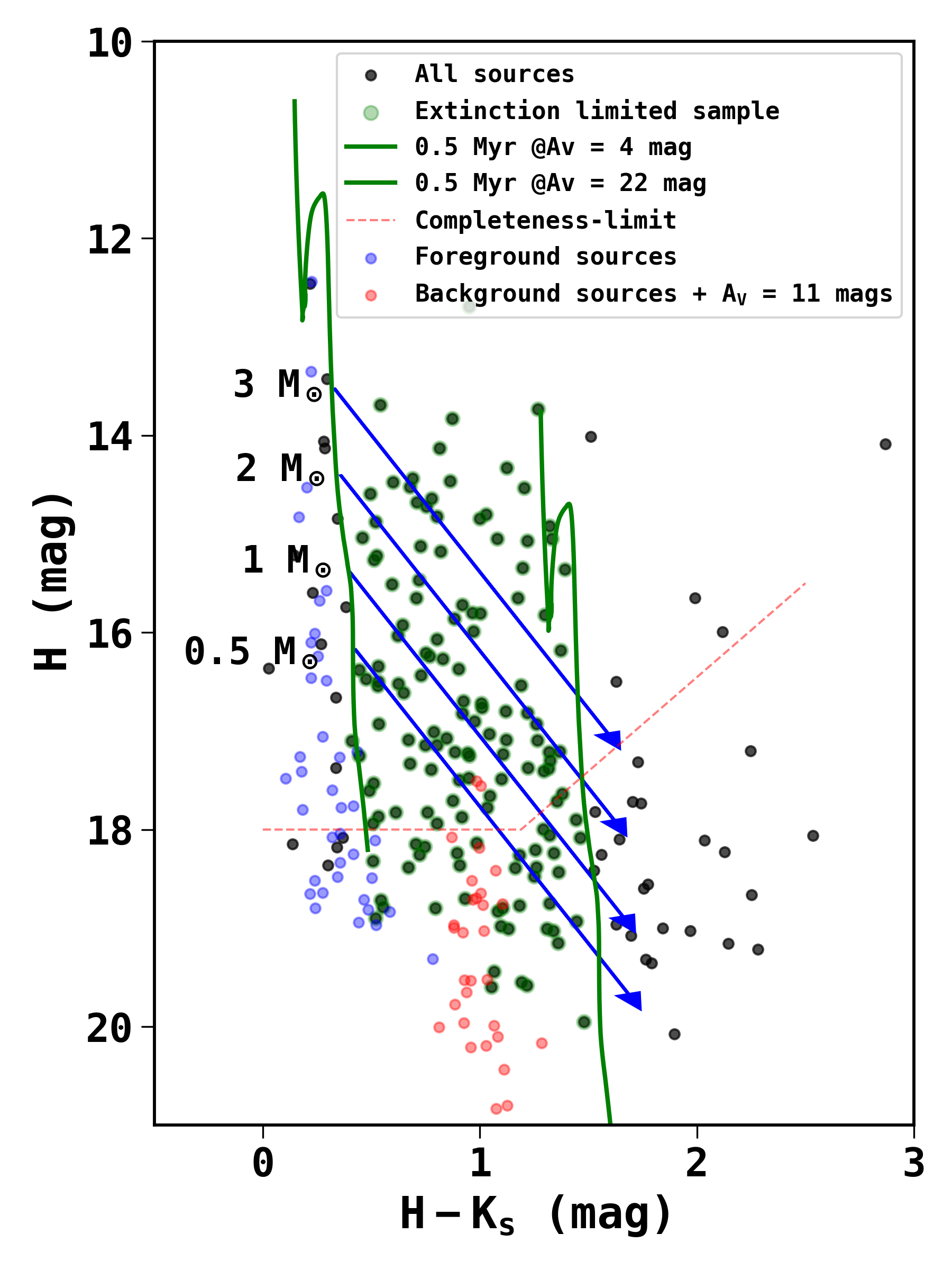}}
\caption{$H$ versus $H-K_s$ color-magnitude diagram of sources in the cluster and control field. The black dots show all the sources observed in the cluster region. The green dots show the sample within \av = 4 to 22 mag. 
The blue dots show the foreground control field sources, and the red dots show the background control field sources, which are reddened to match the median visual extinction of the cluster region, i.e., \av = 11 mag. The MIST isochrones of 0.5 Myr \citep{choi_2016} are reddened by \av = 4 and 22 mag and are shown in green curves. The blue arrows represent the reddening vectors, and the completeness limits of the data are marked by red dashed lines.}
\label{fig_khk}
\end{figure}

As often adopted in young star-forming regions \citep{and11, luh16}, hereafter, we attempted to define a mass-extinction-limited sample of stars to derive further properties of the cluster. The mass-extinction limited sample represents all stars in a given area above a certain mass limit after accounting for the effects of extinction and completeness. 
The primary challenge in obtaining such a sample in young clusters lies in determining the mass limit down to which our data is complete. This determination depends on the age of the cluster and the level of extinction, both of which can be uncertain in young clusters. Unlike open clusters \citep[e.g.][]{sag01,sha06,kum08}, young clusters show variable extinction, which makes it difficult to assign a unique mass to a given source. In order to derive the mass-extinction-limited sample, we use the $H$ versus $H-K_s$ color-magnitude diagram. 
Figure \ref{fig_khk} shows the $H$ versus $H-K_s$ color-magnitude diagram of all the sources. Assuming that the approximate age of the cluster is around 0.5 Myr, we also show a 0.5 Myr MIST isochrone \citep{choi_2016} reddened by \av = 4 and 22 mag in Figure \ref{fig_khk}. 
In the figure, the completeness limit is also shown by the dashed line, while the reddening vectors originating at masses of 3 \Ms, 2 \Ms, 1 \Ms, and 0.5 \Ms~are shown by blue arrows.
To obtain the mass-extinction-limited sample, we choose the visual extinction in the range of 4$-$22 mag, because most of the sources below \av = 4 mag are likely the foreground sources of the field, while only 10\% of the sources lie above \av = 22 mag. 
Applying a high extinction threshold would guarantee a complete sample above a certain mass limit, but it would result in a high minimum mass limit above completeness. We find that \av = 22 mag is a reasonable choice to have a statistically significant number of stars while still reaching fairly low masses above the completeness limit. With \av = 22 mag limit, we find that our sample is better complete above 1 \Ms~and considerably complete down to 0.5 \Ms. 




Field star contamination generally dominates in the low-mass ends of the stellar population. We expect the background and foreground contamination levels in our mass-extinction-limited cluster sample to be minimal. In Figure \ref{fig_khk}, the foreground (blue dots) and background (red dots) population from the Besançon model are also shown. The background populations are reddened to match the median visual extinction of the cluster, i.e., \av = 11 mag. Even with this minimum extinction, we see almost no background population above 0.5 \Ms.

\subsubsection{Mass Function}
\begin{figure}
    \centering
    \includegraphics[width=8cm]{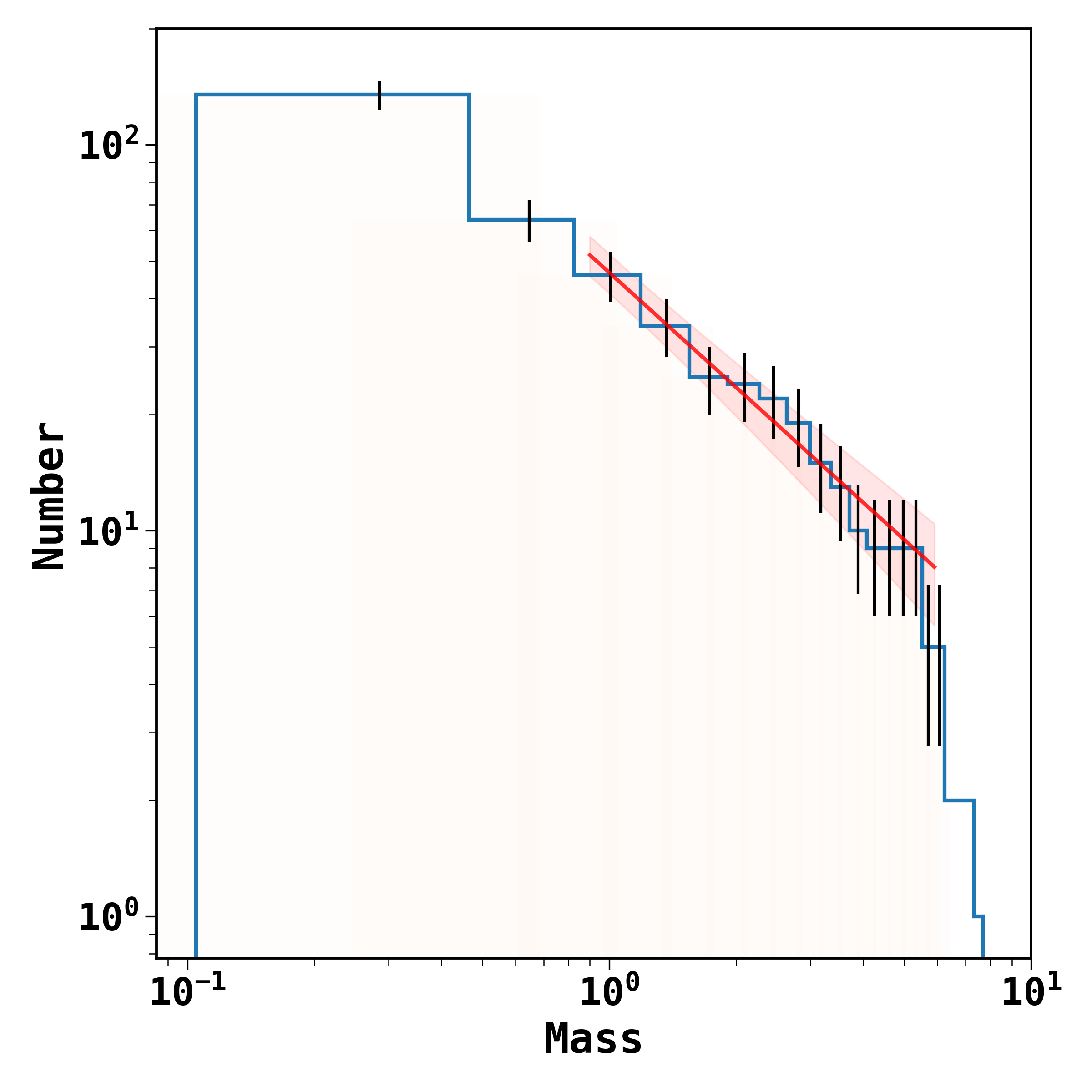}
    \caption{Cumulative initial mass function of the cluster. The red line shows the best-fit power-law mass function with index, $\alpha$ = 1.00 $\pm$ 0.15 for the mass range of 1 to 4 \Ms.}
    \label{fig_imf}
\end{figure}

The stellar IMF describes the mass distribution of the stars at birth in
a stellar system and is fundamental to several astrophysical concepts. Most of the observational studies focusing on the high-mass end (mass $>$ 0.5 \Ms) have found no gross variation of IMF across the Milky Way disk as well as in the local solar neighborhood \citep{sag02, bas10, hop18}, and are in agreement
with Salpeter \citep{sal55} or Kroupa \citep{kro01}-type mass function distribution. At the high-mass end (mass $>$ 0.5 \Ms), the mass function power-law exponent "$\Gamma$" is found to be close to 2.3 in the linear form (i.e., $\frac{dN}{dM} \propto M^{-\Gamma}$) or 1.3 in the logarithm form \citep[i.e. $\frac{dN}{dlogM} \propto M^{-\alpha}$;][]{kro01}, where $\alpha = \Gamma - 1$. 
Studying young clusters, like the one investigated in this work, has the advantage that the dynamical effect of mass segregation will have minimal effect on the shape of the IMF \citep[e.g.][]{pan92, Allison2009b}. In the following, we tried to estimate the IMF of \cluster. 

There are several factors that can affect the IMF shape while dealing with the embedded clusters \citep[e.g.][]{dam21}. The principal factors are the effect of NIR$-$excess and variable extinction in estimating the mass of the stars, low statistics of member stars due to high extinction in getting a robust $\alpha$ value, and contamination at the low-mass end. Since our $J$-band is the least sensitive to the detection of the point sources, to mitigate the effect of NIR-excess sources, we used the $H$-band luminosity because it is less affected by circumstellar matter, compared to other longer wavelengths. To partially mitigate the effect of low statistics, we use the cumulative mass function \cite[e.g.][]{rod12} of the form:

\begin{equation}
N(>M) \propto k M^{-\alpha}
\label{cum}
\end{equation}
Figure \ref{fig_imf} shows the cumulative mass function, $N(>M)$ of the cluster, where $N$ is the number of sources with mass larger than $M$ and the error bars represent the Poisson noise of $\sqrt{N}$.
Using a weighted least-square fit, we find $\alpha$ = $0.95 \pm 0.12$ for the mass range of 0.5 to 4 \Ms. By excluding the points lower than 1 \Ms~in the fitting, to avoid any possible bias that may be introduced by the completeness correction and/or in-proper account of field star contamination at the fainter mass end, we find $\alpha$ =  $1.00 \pm 0.15$ for the mass range of 1 to 4 \Ms. The obtained slopes are, though flatter, but in agreement with the Kroupa IMF \citep{kro01} within 3$\sigma$ uncertainty, where $\sigma$ is the error in our measurements. We exclude sources above 4 \Ms~in fitting as the $M-L$ relation at younger ages (e.g. as found with 0.5 Myr MIST isochrone) is nonlinear in the range 4$-$12 \Ms; thus, a star cannot have a unique mass around this mass range. We note that, though our results show a flatter IMF for \cluster, but should be treated with caution due to various uncertainties involved.


Future more sensitive photometric and spectroscopic observations would improve the robustness of our results with better estimation of extinction, contamination, and contribution from infrared excess. Nevertheless, the characterization of young compact clusters, like the one investigated in this work, is a useful exercise for assessing the mass distribution at the very initial stages of cluster formation.

\subsubsection{Star Formation Efficiency and Rate}

The emergence of a bound cluster also depends on the efficiency with which gas is converted into stars, i.e., SFE ($\epsilon$). The SFE is defined as the ratio of the total stellar mass to the total mass of a star-forming region, i.e., stellar mass plus present-day gas mass.  We estimate the total gaseous mass ($M\rm{_{gas}}$) present in the cluster within its radius. We used the $\it{Herschel}$ molecular hydrogen column density map \citep{Marsh_2017} and determined the total integrated column density ($\Sigma$  $N$(H$_2)$) over the cluster area. We then converted
the total column density to mass using the following equation: $M = \mu m_{\rm{H}} A_{\rm pix} \Sigma N(\rm{H_2})$, 
where $\mu$ is the mean molecular weight taken as 2.8 \citep{Kauffmann_2008}, $m_{\rm{H}}$ is the mass of the hydrogen atom, and $A_{\rm pix}$ is the pixel area of the column density map in cm$^{-2}$ at the distance of the cluster. The gas mass of the cluster region is found to be $\sim$750 $\pm$ 337 \Ms. The uncertainty in the gas mass is around 45\%, which includes the uncertainty in the distance of the cloud, gas-to-dust ratio, and dust opacity index \citep[more details of the procedure can be found in][]{raw23}. 
In order to calculate the mass of the cluster ($M_{\rm{cluster}}$), we integrated the
IMF of the cluster with the Kroupa IMF index, $\Gamma$ = 2.3 \citep{kro01}, within the mass limits of 0.5 to 15 \Ms. Then, we extrapolated down
to 0.08 \Ms~to determine the mass at the lower-mass end, i.e., from 0.5 to 0.08 \Ms, by assuming $\Gamma$ = 1.3 \citep{kro01}. The total stellar mass of the cluster is found to be $\sim$180 $\pm$ 13 \Ms. Using the  $M_{\rm{gas}}$ and $M_{\rm{cluster}}$, we calculated the $\epsilon$ to be around 0.19 $\pm$ 0.07 in the cluster region.

The SFR describes the rate at which the gas in a cloud is converting into stars. 
The SFR can be estimated as, SFR = $M_{\rm{cluster}}/t\rm{_{sf}}$, where  $t\rm{_{sf}}$ is
the star formation timescale.    
Assuming 0.5 Myr as the star formation timescale of the cluster, we obtained the SFR in the cluster region to be around 360 $\pm$ 26 \Ms~Myr$^{-1}$.  
The projected area of the cluster region over which we estimated the cloud mass is calculated as $\pi r^2$ and is found to be $\sim$3.14 $\pm$ 0.57 pc$^{2}$. Here, $r$ = 1 pc is the radius of the cluster region. Normalizing the derived SFR by the cloud area, we got SFR per unit surface area, $\rm \Sigma_{SFR}$ as $\sim$114.6 $\pm$ 22.2 \Ms~Myr$^{-1}$  pc$^{-2}$, whereas the gas mass surface density, $\rm \Sigma_{gas}$ is $\sim$240 $\pm$ 115 \Ms~pc$^{-2}$. 

\citet{kru12} argued that since different clouds can be at different evolutionary stages, therefore normalizing the $\rm \Sigma_{gas}$ with the freefall timescale ($t_{\rm{ff}}$) would give a better correlation with the $\rm \Sigma_{SFR}$. A better correlation of $\rm \Sigma_{SFR}$ with $\Sigma_{\rm{gas}}/t_{\rm{ff}}$ has been found in some studies \citep{kru12, Lee_2016, pokhrel_2021}, and the relation is called as the volumetric star-formation relation. The general form of the relation is expressed as follows \citep{kru12}


\begin{equation}
\Sigma_{\rm{SFR}} = \epsilon_{\rm{ff}} \frac{\Sigma_{\rm{gas}}}{t_{\rm{ff}}},
\end{equation}
where $\epsilon_{\rm{ff}}$ is the star formation rate per freefall time $t_{\rm{ff}}$. \citet{Lee_2016} defined the quantity $\epsilon_{\rm{ff}} = \epsilon~\times~t_{\rm{ff}}/t_{\rm{sf}}$, where $\epsilon$ is the SFE of the cluster.
To test this star-formation relation, we estimate the freefall timescale of the cluster using the following relation

\begin{equation}
t_{\rm{ff}} = \left( \frac{3~\pi}{32~ G \mu m_{\rm{H}} n_{\rm{H_2}}} \right)^{1/2},
\label{eq:tff}
\end{equation}
where $n_{\rm{H_2}}$ is the gas volume number density of the cluster region, and $G$ is the Gravitational constant. The $n_{\rm{H_2}}$ is calculated as $M_{\rm{gas}}/(4/3)\pi r^3 \mu m_{\rm{H}}$, which is around 2637 $\pm$ 1318 \cmq. Using equation \ref{eq:tff}, we calculate $t_{\rm{ff}}$ to be around $\sim$0.60 $\pm$ 0.15 Myr and from $t_{\rm{ff}}$, $t_{\rm{sf}}$ = 0.5 Myr, and $\epsilon$ = 0.19 of the cluster region,  we determined $\epsilon_{\rm{ff}}$ to be around 0.23 $\pm$ 0.10. 

\subsubsection{Possibility of FSR 655 Emerging as a Massive Cluster}


From the stellar population and gas content of the \cluster~region, we estimate an SFE of $\sim$19\% and SFR of $\sim$360 \Ms~Myr$^{-1}$ for the cluster. \cite{raw23, raw24a} found that the cluster is located in a massive clump, which is situated at the filamentary hub of the cloud, and the filaments are inflowing cold gaseous matter at a rate of $\sim$675 \Ms~Myr$^{-1}$ toward the hub. Moreover, \cite{raw24b} found that in the central region of the clump, virial analysis indicates that, at present, the magnetic field and turbulence are not sufficient enough to prevent the collapse of the central clump region. So, we hypothesize that,
if the star formation continues in the clump with the current rate for another 2 Myr along with the continuous mass supply through the filaments, then a cluster of total stellar mass $\sim$1000 \Ms~ is expected to emerge at the hub of \cloud.

\section{Summary and Concluding Remarks}
\label{summary_con}
In order to better understand the formation of star clusters in GMCs, in this work, we have studied a young cluster, \cluster~located at the hub of the \cloud~cloud. To study the cluster properties, we used the TANSPEC NIR camera mounted on the 3.6 m DOT. With TANSPEC, we probe the cluster down to 5$\sigma$ limiting magnitudes of 20.5, 20.1, and 18.6 mag in the $J$, $H$, and $K_s$ bands, respectively. 

The cluster shows differential extinction with a mean visual extinction of $\sim$11 mag, whereas the foreground visual extinction in the direction of the cluster is around 4 mag. 
The age of the cluster derived by matching the KLF of the cluster members with the KLFs of the synthetic clusters is found to be around 0.5 Myr. Using the $JHK_s$ and $HK_s[4.5]$ CC diagram, we find the disk fraction around 38\% $\pm$ 6\% and 57\% $\pm$ 8\%, respectively. 

The mass distribution function of the cluster members agrees with the Kroupa IMF within a 2$\sigma$ uncertainty, with an $\alpha$ index value of around 1.00 $\pm$ 0.15 for the mass range of 1 to 4 \Ms. Using the Kroupa IMF, we determined the present-day total stellar mass of the \cluster~cluster to be $\sim$180 $\pm$ 13 \Ms. The gas mass of the  cluster is around 750 $\pm$ 337 \Ms, which gives an SFE of $\sim$19\% $\pm$ 7\% and an SFR of $\sim$360 $\pm$ 26 \Ms~Myr$^{-1}$.

The robust estimation of all these parameters certainly needed
deep high-resolution photometric observations down to the brown-dwarf regime. Nonetheless, taking these results at face value and assuming a constant SFR for a time span of 2 Myr, we find that the cluster has the potential to grow further to become a 1000 \Ms~cluster. In previous studies on \cloud~\citep{raw23}, the converging gas flows toward the hub of the cloud have been found which are supplying the necessary material for star and star cluster formation at the hub. 
Given the fact that the cluster is located near the geometric center of the cloud, whose mass is $\sim$10$^5$ \Ms, and from the evidence of gas in-fall onto the region at a high rate via large-scale 
filamentary flows, it is not unreasonable to think that the cluster will increase in mass in the future and may emerge as a massive cluster. Moreover, simulations suggest an accelerated pace of star formation in molecular clouds with SFR $\propto$ t$^2$ due to global hierarchical and runway collapse of molecular gas up to a few Myr since the beginning of the star formation \citep[e.g.][]{cal18,vaz19}. So, the possibility of \cluster~becoming a more massive and richer cluster seems to be viable. Future NIR proper motions and radial velocity observations, combined with detailed simulations of stellar and gas motions simultaneously, would be highly desirable to test the above hypothesis for the studied cluster as well as for embedded clusters in general.  

\section*{Acknowledgments}
We thank the anonymous referee for the constructive comments and suggestions that helped to improve the paper.
The research work at the Physical Research Laboratory is funded by the Department of Space, Government of India. This work makes use of the data observed from the 3.6m DOT at Devasthal, Nainital, India, and we thank the staff of the telescope operation team. We also acknowledge the TIFR-ARIES Near Infrared Spectrometer mounted on 3.6m DOT using which we have made NIR observations. This work also uses the data based on observations made with the Spitzer Space Telescope, which was operated by the Jet Propulsion Laboratory, California Institute of Technology under a contract with NASA. D.K.O. acknowledges the support of the Department of Atomic Energy, Government of India, under Project Identification No. RTI 4002.

%

\vspace{5mm}
\facilities{ARIES:DOT}


\software{astropy \citep{2013A&A...558A..33A,2018AJ....156..123A},  
         Source Extractor \citep{Bertin_1996}, Besançon population synthesis model \citep{Robin_2004}, SPISEA \citep{Hosek_2020}
          }



\appendix

\section{NIR-excess Sources }
\label{sec_cc}
\begin{figure*}
\centering{
\includegraphics[width=14.0cm]{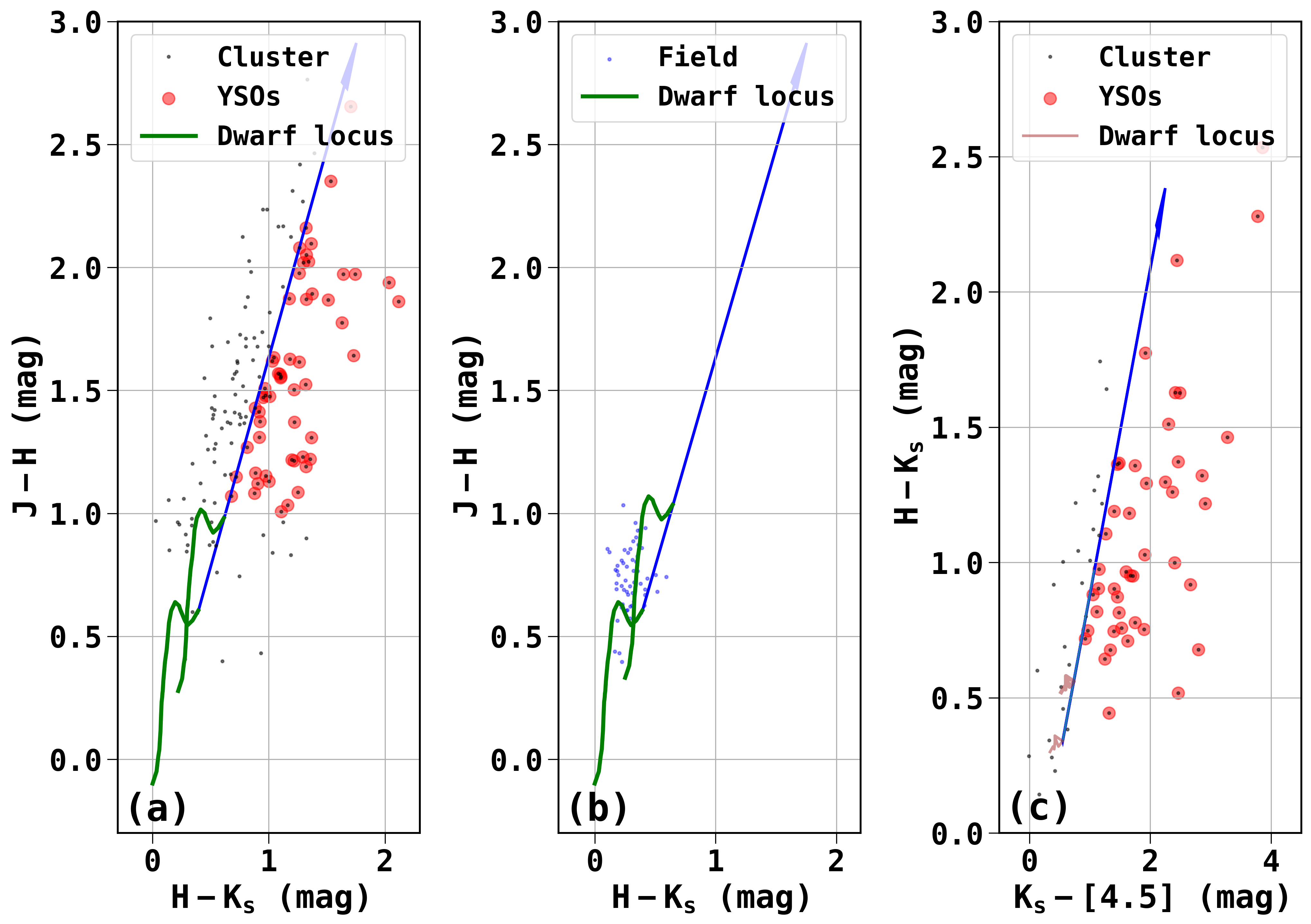}}

\caption{($J-H$, $H-K_s$) CC diagram for the (a) cluster region and (b) for the modeled control field region. The green curves are the intrinsic dwarf locus from \citet{bes88}. The blue dots in panel-b show the modeled field population. (c) The ($H-K_s$,  K$-$[4.5])  CC diagram for the cluster region. The brown curves are the intrinsic dwarf locus of late M-type dwarfs \citep{pat06}. In panel-a and -c, the black dots show all the sources observed toward the cluster, and the red dots show the YSOs identified in the cluster, based on their NIR$-$excess in the $JHK_s$ and $HK_s[4.5]$ CC diagrams, respectively. In all the plots,
the blue line represents the reddening vector drawn from the location of the M6 dwarf.}
\label{fig_irac}
\end{figure*}

For deriving cluster properties, the identification of cluster members is crucial. The near and mid-infrared CC diagrams are useful tools to identify the cluster members having  NIR$-$excess emission due to circumstellar disks from young stars. However, other dusty objects along the line of
sight may also appear as NIR$-$excess sources in the CC diagram. Without proper motion or radial velocity information, it is difficult to separate the member sources from the reddened field sources. One possible way
to separate out the members from the field sources is to compare the CC diagram of the cluster with that of the field
sources of the same area and photometric depth. Thus, we made the CC diagrams for the cluster as well as population synthesis model stars and did a comparative
analysis of the distribution of the sources.  Figure \ref{fig_irac}(a) and (b) show $J-H$ versus $H-K_s$ CC ($JHK_s$ CC) diagrams of the cluster as well as model field sources, respectively. In both diagrams, the main-sequence dwarfs' locus is shown by a green curve, and the reddening vector from the location of the  M6 
dwarf is shown by a blue arrow. In the NIR CC diagram, sources right to the M6 dwarf reddening vector are, in general, considered as pre-main-sequence (PMS) sources with NIR$-$excess \citep{lada92, lada95, hai01}. 

As can be seen in Figure \ref{fig_irac}, compared to the cluster region, the NIR$-$excess zone of the field population is mostly devoid of sources, implying the presence of true NIR$-$excess sources in the cluster region. From the figure, it can also be noticed that most of the control field sources are distributed in the CC space of $J-H$ $<$ 1.0 mag and $H-K_s$ $<$ 0.4 mag. A similar distribution with $J-H$ color less than 1.2 mag can also be seen for a group of sources in the cluster CC diagram. This group seems to be separated from the group of reddened cluster sources in the $J-H$ and $H-K_s$ color space.
A comparison of CC diagrams leads us to suggest that the former group of sources in the cluster region is likely the field population along the line of sight. Figure \ref{fig_irac}(a) and (b) also
show the location of the dwarf locus reddened by \av = 4 mag, which fairly matches with the distribution of control field sources and also the likely field population of the cluster region, implying that foreground extinction in front of cluster hosting cloud is around 4 mag. Comparing the CC diagrams, we selected sources with ($J-H$) color greater than 1.0 mag as NIR$-$excess sources.

It is well known that circumstellar emission from young stars dominates at longer wavelengths, where the spectral energy distribution (SED) significantly deviates from the pure photospheric emission. Thus, by incorporating the $\it{Spitzer}$ longer wavebands' data into the analysis, a more accurate census of the fraction of stars still surrounded by circumstellar material (i.e., optically thick accretion disks) can be obtained. We thus used the $H-K_s$ versus $K_s-[4.5]$ CC ($HK_s[4.5]$ CC) diagram to identify extra NIR$-$excess sources \citep[e.g.,][]{sam14}, which is shown in Figure \ref{fig_irac}c. 
Similar to the $JHK_s$ CC diagram, 
we selected NIR$-$excess sources whose 
($H-K_s$) color is greater than 0.5 mag and located right to the 
reddening vector drawn from the M6 dwarf star.

\begin{figure}
    \centering
    \includegraphics[width=8.0cm]{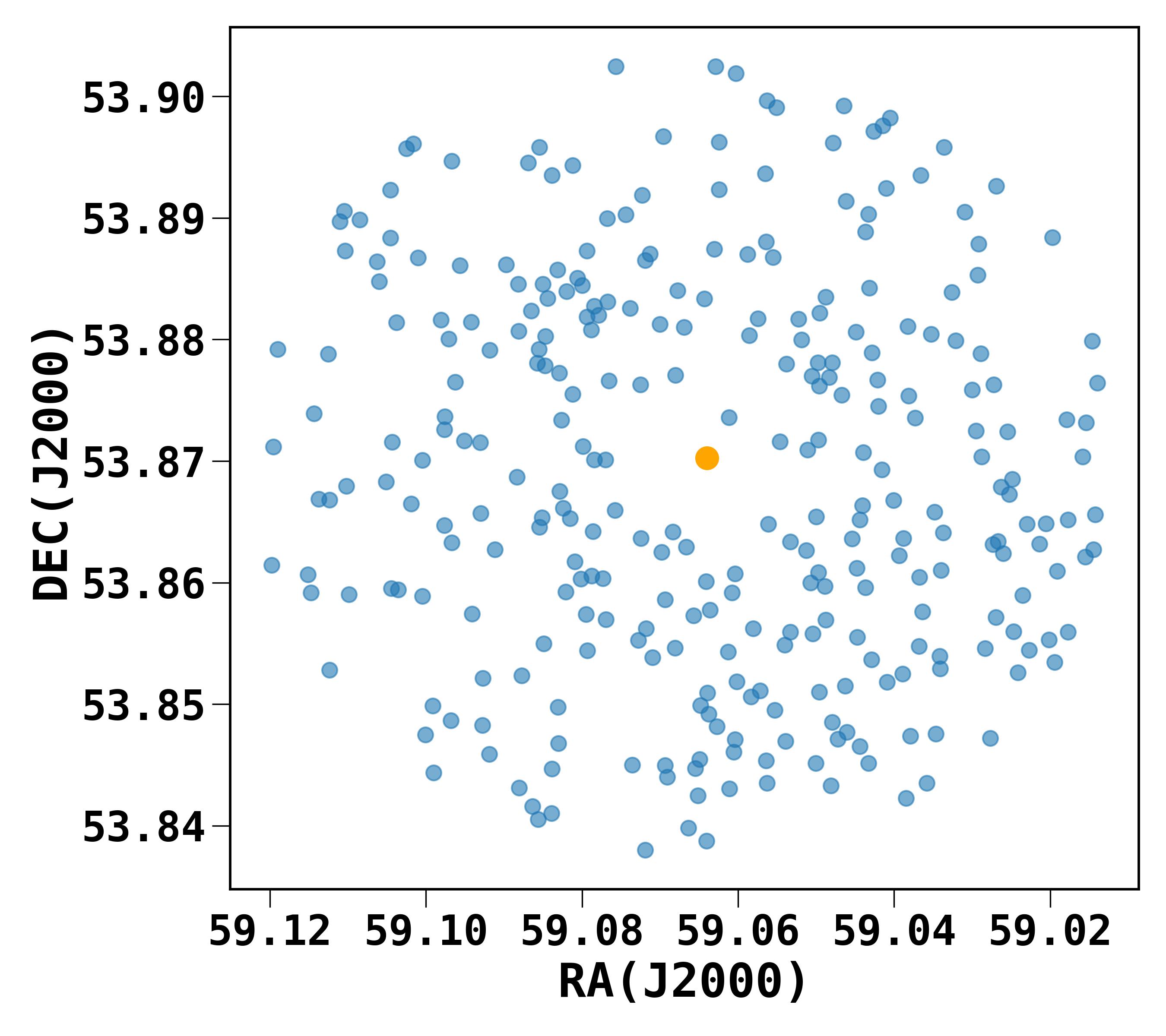}
    \caption{Spatial distribution of the sources visible in the 4.5\mum band, in the cluster direction within a 2 arcmin radius from the cluster center. The location of the central massive YSO is marked by a yellow dot.}
    \label{fig_hole}
\end{figure}
 
 In summary, with the above approaches, we identified 56 and 47 NIR$-$excess sources from the $JHK_s$ and $HK_s[4.5]$ CC  diagrams, respectively. Including common sources, in total,  we identified 82 disk-bearing sources in the cluster region. 

\section{Disk Fraction}

\label{disk_frac}
The disk fraction, which is the frequency of stars with disks within a young cluster, has been widely studied for various star-forming clusters in the solar neighborhood. In general, it has been found that the disk fraction decreases exponentially with the age of the cluster, and the typical lifetime of an optically thick circumstellar disk is around 2$-$3 Myr \citep {hai01}. 

Using the $JHK_s$ and $HK_s[4.5]$ CC diagrams discussed in Appendix \ref{sec_cc}, we estimate the disk fraction of \cluster~to be around 47\% $\pm$ 7\% and 70\% $\pm$ 8\%, respectively, where the errors are due to Poisson statistics. However, if we include the photometric error of the cluster members and select only those sources that have excess 1$\sigma$ (where $\sigma$ is the color error) above the reddening vector, the disk fraction changes to $\sim$38\% $\pm$ 6\% and $\sim$57\% $\pm$ 8\%, respectively. To further confirm the disk-bearing cluster members, we determine the $Q$ parameter, $Q = (J-H) - 1.7 \times (H-K_s)$, which gives the deviation from the reddening vector in the $JHK_s$ CC diagram \citep{Comeron_2005, messi_2012}, following the \cite{rie85} extinction law. 
Following the criteria of \cite{Comeron_2005}, i.e., a star having a $Q$ value of less than $-$0.10 is an NIR-excess source, we estimated the $JHK_s$ disk fraction of \cluster~to be around 43\%. We also find that using a higher alpha value of 1.9 in the extinction law (discussed in Section \ref{extin}), the disk fraction of \cluster~based on $Q$ value estimation changes by only 2\%. A similar analysis of $Q$ value for $HK_s[4.5]$ CC-based disk fraction shows comparable results within 1$\sigma$ uncertainty.  

Comparing the disk fractions of \cluster~with those of the NGC 2024 cluster, which is of similar age $\sim$0.3 Myr \citep{hai00}, we find that the $JHK_s$ and $HK_s[4.5]$ disk fractions of \cluster~are comparable to the $JHK_s$ and $JHK_sL$ disk fractions of NGC 2024 (i.e., $\sim$58\% and $\sim$86\%, respectively) within the limits of uncertainty.
However, we want to point out that the $HK_s[4.5]$ CC-based disk fraction estimated for \cluster, 
is likely a lower limit. This is due to the presence of a high infrared diffuse background in the vicinity of the cluster center at 3.6 $\mu$m and 4.5 $\mu$m, which can potentially affect the detection of the faint point sources in these bands. This can be readily seen in Figure \ref{fig_hole}, as a lack of point sources in the vicinity of the central massive star, shown by a yellow dot, compared to the overall distribution of point sources in the area. 
Deeper and high-contrast observations are required to determine the true $JHK_sL$ or $HK_s[4.5]$ based disk fraction of the cluster. 

Furthermore, it has been suggested that the gas and dust in the disk are affected by the stellar radiation of the host stars, thus, the disk fraction also depends on stellar mass. For example, larger disk fractions among lower-mass stars, compared to massive stars, have been found both in simulations \citep{Johnstone_1998, Hollen_2000, Pfalzner_2006, Pfalzner_2024} and observations \citep{Balog_2007, KK_2009, Stolte_2010, Yasui_2014, Ribas_2015, Damian_2023}. It is thus important to estimate the disk fraction in a limited mass range. In this line, \cite{fang12} estimated inner-disk fraction based on $H$, $K_s$, $3.6$, and $4.5$ $\mu$m data for a number of nearby clusters with stellar members massive 
than 0.5 \Ms~and found the dependence of disk fraction ($f\rm{_{disk}}$) on age as $f\rm{_{disk}} = e^{-t/2.3}$, where $t$ is the age in Myr. They
also found that for clusters having a higher number of OB stars, the disk dispersal is faster compared to the moderate number of OB stars. We thus estimated disk fraction using the mass-extinction limited sample, which is 
fairly complete, down to 0.5 \Ms. Doing so, we find the $JHK_s$ CC and $HK_s[4.5]$ CC disk fraction to be around 45\% $\pm$ 7\% and 65\% $\pm$ 8\%. This may be a lower limit considering that our data is not fully complete down to 0.5 \Ms.

Comparing the $HK_s[4.5]$ CC disk fraction of \cluster~with the samples of \cite{fang12} (see their Figure 16), we find that the likely age of the cluster is not more than a Myr. To our knowledge, no disk fraction in the literature has been estimated for cluster members of mass above 0.5 \Ms~using only $JHK_s$ data, so a direct comparison of the $JHK_s$ disk fraction with other clusters is not possible. However, in general, it is comparable to the disk fraction ($\sim$50\%$-$60\%) of nearby clusters of age 0.5$-$1 Myr such as NGC 2024 and ONC \citep{hai00,lada00}, for which disk fraction has been estimated for member stars down to 0.1 \Ms.


\bibliography{myref}{}
\bibliographystyle{aasjournal}



\end{document}